\documentclass[a4paper,11pt]{article}
\usepackage{jheppub} 
\usepackage{lineno}
\usepackage{cancel}
\usepackage{physics}
\usepackage{xcolor}
\usepackage{multirow}
\usepackage{comment}
\usepackage{array}
\usepackage{subcaption} 
\usepackage{float}
\usepackage{textcase}
\usepackage{verbatim} 

\title{\boldmath Effect of Cosmic Neutrino Background on the Dark Matter Self-interaction via Neutrino force}

\makeatletter
\gdef\@fpheader{%
  \normalfont\small
  \href{https://doi.org/10.1007/JHEP05(2026)028}{%
    \textcolor{black}{\nolinkurl{https://doi.org/10.1007/JHEP05(2026)028}}%
  }%
}
\gdef\@journal{jhep}
\makeatother

\author[a]{Pawin Ittisamai,}
\author[b]{Chakrit Pongkitivanichkul,}
\author[a,c]{Muhammaddaniya Sutwilai,}
\author[b]{Nakorn Thongyoi,}
\author[d]{and Patipan Uttayarat}
\affiliation[a]{Department of Physics, Faculty of Science, Chulalongkorn University, 254 Phayathai Road, Pathumwan, Bangkok, Thailand}
\affiliation[b]{Khon Kaen Particle Physics and Cosmology Theory Group (KKPaCT), Department of Physics,
 Faculty of Science, Khon Kaen University, 123 Mitraphap Rd, Khon Kaen 40002, Thailand
}
\affiliation[c]{Abdus Salam International Centre for Theoretical Physics (ICTP), Strada Costiera 11, 34151,
Trieste, Italy}
\affiliation[d]{Theoretical High-Energy Physics and Astrophysics Research Unit (ThEPA), Department of Physics, Srinakharinwirot University, 114 Sukhumvit 23 Rd., Wattana, Bangkok 10110, Thailand}

\emailAdd{pawin.i@chula.ac.th}
\emailAdd{chakpo@kku.ac.th}
\emailAdd{muhammaddaniya.sutwilai@gmail.com}
\emailAdd{nakorn.thongyoi@gmail.com}
\emailAdd{patipan@g.swu.ac.th}

\abstract{Neutrino–pair exchange induces a neutrino force that can drive dark matter (DM) self-interactions and impact small-scale structure formation. In the presence of the cosmic neutrino background (C$\nu$B), this force can be modified, with important consequences for DM phenomenology. We study the effect of the C$\nu$B on neutrino forces, generated by the scalar and pseudoscalar interactions. We explore the significance of the background neutrino force on the scalar DM-neutrino portal model, including DM self-scattering and annihilation. Our results show that the interplay between attractive vacuum potential and repulsive background potential leads to a screening effect that varies across DM mass ($m_\chi$) regimes, strongly affecting DM self-scattering in the DM mass $m_\nu \lesssim m_\chi \lesssim  T_{C \nu B}$. Meanwhile, for DM annihilation, the screening completely vanishes the Sommerfeld Enhancement induced by the neutrino force. Overall, the C$\nu$B substantially reshapes the viable coupling range for DM self-interactions while remaining compatible with current constraints, offering a pathway to small-scale structure problems.
}

\begin{document}
\maketitle
\flushbottom

\section{Introduction}
\label{sec:intro}

Dark matter (DM) is the hypothetical matter, introduced to solve several problems in cosmology and astrophysics~\cite{RevModPhys.90.045002, Cirelli:2024ssz}. Although there are several pieces of evidence indicating the existence of DM~\cite{Rubin:1970zza, inproceedings, Planck:2018vyg}, its fundamental nature has not been confirmed. Its existence is implied solely by its gravitational effects~\cite{Clowe:2006eq}, which resolve the discrepancies between theoretical predictions and astronomical observations. Therefore, several models~\cite{ParticleDataGroup:2024cfk, Bertone:2004pz, Feng:2010gw, Arbey_2021, Arg_elles_2023, Matos:2023usa} with explicit assumptions of DM have been studied to confirm the gravitationally inferred observables such as stars' motion in the galaxies and structure formation. Lambda-Cold Dark Matter ($\Lambda \text{CDM}$) is one of the best descriptions used to explain standard cosmology, including the cosmic microwave background (CMB), large-scale structure, and the accelerating expansion of the universe ~\cite{Sahni:1999gb, Planck:2018vyg, Dodelson:2003ft, Baumann:2022mni}. Although cold dark matter is successful at explaining the large-scale structure of the Universe, discrepancies arise at a small scale, such as the `core-cusp', `missing satellite problem', and `too big to fail' problems~\cite{Moore:1999nt, Moore:1999gc, Boylan-Kolchin:2011qkt,adhikari2022astrophysicaltestsdarkmatter, Tulin_2018, DelPopolo:2016emo}. These discrepancies emerge when comparing observations with theoretical predictions from DM-only simulations ~\cite{Navarro:1995iw, Navarro:2008kc,2014MNRAS.445.3512S, Klypin:1999uc}. These problems might be alleviated by the baryonic process~\cite{Pontzen:2014lma, Navarro:1996bv, Governato:2009bg} or the warm dark matter~\cite{Lovell:2011rd, Lovell:2013ola, Maccio:2012qf}. Another promising alternative is to replace the assumption of collisionless CDM with that of Self-Interacting Dark Matter (SIDM)~\cite{PhysRevLett.84.3760, Tulin_2018, adhikari2022astrophysicaltestsdarkmatter}. This new non-gravitational force could compete with gravity and affect small-scale structure. The dynamics of SIDM have been explored in the literature, see e.g.~\cite{Vogelsberger:2012sa, Tulin:2013teo, Tulin_2018, Fichet:2017bng, Costantino:2019ixl, Costantino:2020bei}. The light mediator is often introduced to account for the long-range behavior of the interaction potential. However, this mediator typically belongs to the dark sector, presenting a challenge in probing it through the Standard Model (SM) portal.

A well-motivated SM-mediated dark-matter self-interaction arises from the exchange of a pair of light neutrinos. Assuming the contact interaction, this generates a long-range effective potential of the form $1/r^5$ at separations shorter than the inverse neutrino mass ~\cite{Feinberg:1968zz, Grifols_1996, Hsu:1992tg, Costantino:2020bei}. The DM-neutrino interaction is interesting in several aspects. It naturally leads to neutrino-portal scenarios, which are highly testable through constraints from SM decays~\cite{Gonzalez-Macias:2016vxy, Blennow:2019fhy, Orlofsky_2021, Zhang_2024}. Moreover, on the observational side, these interactions feature a rich cosmological signal and phenomenology~\cite{Wilkinson:2014ksa, Bertoni:2014mva, Olivares-DelCampo:2017feq, Ko:2014bka, Batell:2017rol, Boehm:2017dze, Becker:2018rve}. For SIDM, the neutrino as a force carrier needs no further assumption about the portal to the SM. Moreover, this requires only SM neutrinos as light mediators and can modify the thermal history of DM, which opens up the scenario of warm dark matter~\cite{Bertoni:2014mva}. For DM annihilation, the neutrino force can play a non-negligible role in Sommerfeld enhancement~\cite{Coy_2022}. More importantly, it was shown that the neutrino force between DM is strong enough to impact small-scale structure formation~\cite{Orlofsky_2021, Zhang_2024} with the emphasis on DM mass around MeV to a few GeV scale. In the presence of the neutrino background, this interaction can be significantly modified. This results in a modification of the effective potential, which depends on the type of interaction and the background distribution profile (see refs.~\cite{Horowitz:1993kw, Ferrer:1998ju, VanTilburg:2024xib, Barbosa:2024pkl, Ghosh_2023, Ghosh:2024qai,ghosh2024neutrinoforcelengthscales} and also \cite{Day:2023mkb,Gan:2025nlu} for wave-like dark matter). It has been shown that the background correction on the quantum force between DM is significant in modifying DM annihilation~\cite{Ferrante:2025lbs}, where the Sommerfeld enhancement is investigated for quadratic effective operators of both bosonic and fermionic mediators. This also includes thermal background corrections, which result in both the DM mass and temperature-dependent enhancement and suppression effects.
The effect of the cosmic neutrino background (C$\nu$B) might also enhance or suppress DM annihilation and self-scattering. Notably, ref.~\cite{Orlofsky_2021} focused on asymmetric fermionic DM, for which the neutrino-mediated force is repulsive and sub-keV masses are excluded. By contrast, scalar DM admits an attractive potential and allows for lower-mass regimes, motivating a dedicated study of scalar-DM self-scatterings via the neutrino force.

Building on recent progress in exploring DM-neutrino interactions, we extend the study to incorporate the effects of the C$\nu$B on the effective potential and its implications for DM self-interactions. In particular, the role of background potential from the C$\nu$B, modifying the potential (see ref.~\cite{Ghosh_2023}), on the DM phenomenology has not been fully addressed in the literature. We analyze how the C$\nu$B modifies the long-range neutrino-mediated potential for scalar DM and assess the resulting implications for observables, including the self-scattering cross section and Sommerfeld enhancement in DM annihilation.

In this work, we investigate the role of neutrino forces in scalar DM self-interactions, focusing on the impact of C$\nu$B on DM phenomenology. The work is organized as follows. In section~\ref{sec: neutrino force}, we set up the general formalism to calculate the neutrino vacuum and background potential.  In section~\ref{sec: phenomenology}, we investigate the impact of C$\nu$B on DM phenomenology, including the self-scattering and annihilation. The allowed region of parameter space is derived in this part. Following this analysis on DM phenomenology, we discuss the scalar dark matter model-building with a focus on generating the given DM-neutrino effective operator. Additionally, the laboratory constraints from the possible UV channel are discussed. Our main conclusions are summarized in section~\ref{sec: conclusion}. Technical details are provided in the appendices.

\section{Neutrino force in the dark matter self-interaction}
\label{sec: neutrino force}
In this section, we calculate the potential associated with DM self-interaction via neutrino force, considering both vacuum and background contributions to the potential. First, we outline the formalism in section~\ref{subsec:formalism} to establish the theoretical framework for neutrino-based DM interactions and potential calculations. We then derive the vacuum potential in section~\ref{subsec: vacuum potential contact} for both the scalar and pseudoscalar types of interaction. Next, we review the calculation in ref.~\cite{Ghosh_2023} and study the background potential in section~\ref{subsec: bkg potential}. Finally, we examine the implication of the C$\nu$B effect on the DM self-interaction in section~\ref{subsec: implication of bkg on DM scattering}.

\subsection{Formalism}
\label{subsec:formalism}
\begin{figure}[h]
\centering
\includegraphics[width=0.40\textwidth]{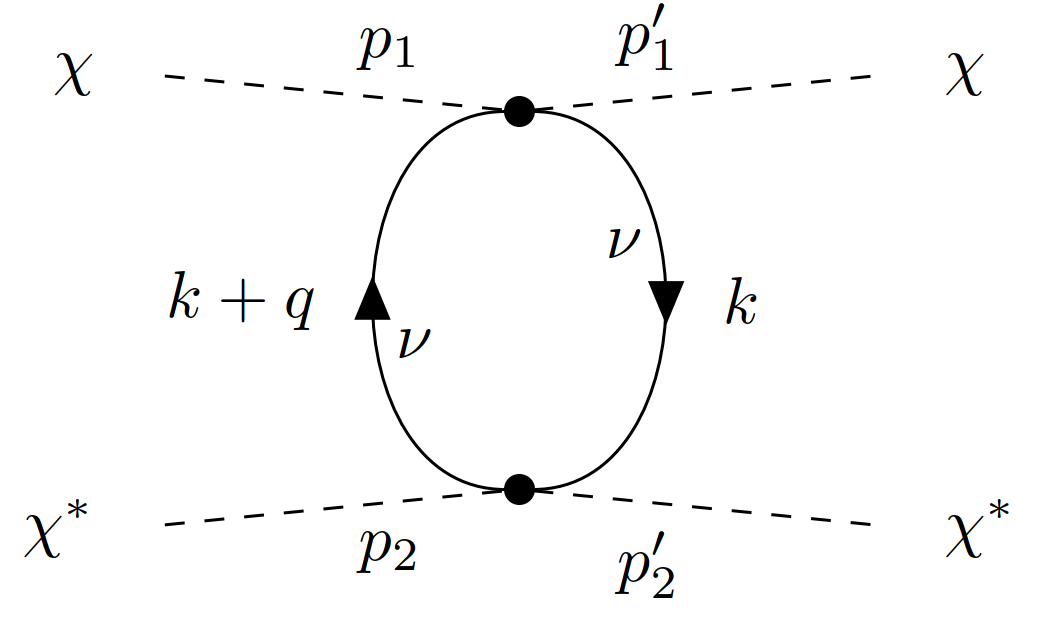}
\caption[Short-form caption]{Scalar DM self-scattering via double neutrino exchange.}
\label{fig: vac vs bkg diagram}
\end{figure}

In this work, we consider quadratic interactions between a massive Majorana neutrino $\nu$ and a CP-even complex scalar DM $\chi$ via the scalar-type and pseudoscalar-type interactions 
\begin{equation}
    \mathcal{L} \supset \frac{G_{s}}{\sqrt{2}} \abs{\chi}^2 \Bar{\nu^c} \nu + \frac{G_{p}}{\sqrt{2}} \abs{\chi}^2 \Bar{\nu^c} i \gamma^5 \nu ,
    \label{Effective DM Interaction}
\end{equation}
where $\nu = \nu_L + \nu_L^c$. Note that the pseudoscalar-type interaction gives rise to CP violation. These interactions induce a quantum force for DM self-interaction through the exchange of two neutrinos in a fermion loop, as shown in figure~\ref{fig: vac vs bkg diagram}. For the scattering $\chi \chi^* \rightarrow \chi \chi^*$, only the $t$-channel neutrino exchange diagrams contribute. The effective couplings $G_s$ and $G_p$ are obtained by integrating out the heavy field in the UV model. This will be discussed in section~\ref{subsec: UV channel}.

Since DM is non-relativistic (NR), with typical speed $\sim 10^{-3}-10^{-5}$,  the momentum exchange $q \equiv p_1^{'} - p_1$ in the NR limit satisfies $q \ll m_\chi$, where $m_\chi$ is the DM mass. This ensures the validity of a potential description.

The potential is evaluated from the non-relativistic scattering amplitude of the double neutrino exchange diagram by a Fourier transform,
\begin{equation}
    V(\bold{r}) = - \int \frac{d^3\bold{q}}{(2 \pi)^3} e^{i \bold{q} \cdot \bold{r}} \mathcal{A}(\bold{q}),
    \label{fourier transform}
\end{equation}
where $\mathcal{A} = \mathcal{M}/4m_\chi^2$ is the normalized scattering amplitude. In the presence of a neutrino background, the neutrino propagator is modified. Following  ref.~\cite{Ghosh_2023}, the modified neutrino propagator can be written as
\begin{equation}
    S_\nu(k) = (\cancel{k} + m_\nu) \left[ \frac{i}{k^2 - m_\nu^2 + i\epsilon} - 2\pi \delta(k^2 - m_\nu^2) \{ \Theta(k^0) n_+ + \Theta(-k^0) n_- \} \right] \equiv S^0_\nu + S^\text{bkg}_\nu,
    \label{modified propagator}
\end{equation}
where $\Theta(k)$ is the Heaviside step function, $n_+$ and $n_-$ are the number densities of neutrinos and antineutrinos, respectively. Using the modified propagator, the total amplitude of two neutrinos exchanges can be written as the sum of the contribution from vacuum and background as $\mathcal{A}(\textbf{q}) = \mathcal{A}_\text{vac}(\textbf{q}) + \mathcal{A}_\text{bkg}(\textbf{q})$ in which the background contribution comes from the cross term of $S^0_\nu$ and $S^\text{bkg}_\nu$. This can be viewed as replacing one virtual neutrino leg with an on-shell neutrino from the background. The corresponding potential decomposes as
\begin{equation}
    V(\textbf{r}) = V_\text{vac}(\textbf{r}) + V_\text{bkg}(\textbf{r}) ,
\end{equation}
allowing us to compute and interpret vacuum and background effects separately in section~\ref{subsec: vacuum potential contact} and \ref{subsec: bkg potential}.

\subsection{Calculation of the vacuum potential}
\label{subsec: vacuum potential contact}
The effective operators in Eq.~\eqref{Effective DM Interaction} generate the DM self-scattering as shown in figure~\ref{fig: vac vs bkg diagram}. In the NR limit, the potential follows from the Fourier transform of the scattering amplitude; equivalently, it is determined by the discontinuity across the branch cut of this diagram's amplitude. Detailed calculations and general potential forms are provided in Appendix~\ref{appendix: Calculation of the massive neutrino vacuum potential}. The vacuum potential for both interactions is attractive and exhibits the same $1/r^5$ dependence in the short-range limit ($r \ll m_\nu^{-1}$), as follows:
\begin{align}
    V_\text{vac}^s(r \ll m_\nu^{-1}) &= -\frac{3G_s^2}{32 \pi^3 m_\chi^2 r^5}, \\
    V_\text{vac}^p(r \ll m_\nu^{-1}) &= -\frac{3G_p^2}{32 \pi^3 m_\chi^2 r^5}. 
\end{align}

At long range ($r \gg m_\nu^{-1}$), the vacuum potential is exponentially suppressed. In particular, the vacuum potential for scalar and pseudoscalar interactions has different asymptotic form as $V^s_\text{vac} \sim \left (\frac{m_\nu^3}{r^7} \right)^{1/2}e^{-2m_\nu r}$ and $V^p_\text{vac} \sim \left (\frac{m_\nu}{r} \right)^{5/2}e^{-2m_\nu r}$, respectively. But this difference is small in the $r \gg m_\nu^{-1}$ regime.

The qualitative features of the neutrino force from the effective interaction in  Eq.~\eqref{Effective DM Interaction} -- attractiveness and the $r$-dependence -- follow from the optical theorem and dimensional analysis (see ref.~\cite{Brax_2018}). In particular, the sign of the potential is fixed by the sign of the discontinuity across the branch cut, whereas the $1/r^5$ short-distance behavior is dictated by the dimensional analysis. 

\newpage

\subsection{Calculation of the background potential}
\label{subsec: bkg potential}
We now derive the background potential of the scalar and pseudoscalar type interactions following ref.~\cite{Ghosh_2023, Ghosh:2024qai} for the contact interaction. In the presence of a neutrino background, the corresponding background contributions of the matrix elements of both interactions are
\begin{align}
    i\mathcal{M}_\text{bkg}^s &= G_s^2 \int \frac{d^4 k}{(2 \pi)^4} {\rm Tr}\left[ S^0_\nu(k)S^\text{bkg}_\nu(k+q) + S^\text{bkg}_\nu(k)S^0_\nu(k+q) \right], \\
    i\mathcal{M}_\text{bkg}^p  &= -G_p^2 \int \frac{d^4 k}{(2 \pi)^4} {\rm Tr}\left[ \gamma^5 S^0_\nu(k) \gamma^5 S^\text{bkg}_\nu(k+q) + \gamma^5 S^\text{bkg}_\nu(k) \gamma^5 S^0_\nu(k+q) \right].
\end{align}
The background potential is obtained from the NR amplitude via Fourier transform. Since we are interested in the background effect on small-scale structures, we assume that the C$\nu$B is the most dominant due to its density and its presence in a long thermalization time. Due to the C$\nu$B isotropy, we can simplify the 3D-momentum integral to be in the form of the magnitude of the spatial momentum $\kappa$. Details are provided in Appendix~\ref{appendix: Calculation of the massive neutrino background potential}. The resulting isotropic background potentials are
\begin{equation}
    V^s_\text{bkg} = +\frac{G_s^2}{16\pi^3 m^2_\chi r^4} \int_0^{\infty} d\kappa\; \kappa \frac{n_+(\kappa) + n_-(\kappa)}{\sqrt{\kappa^2 + m_\nu^2}}  \Big\{ \left[1- 2\left(\kappa^2+m_\nu^2\right)r^2\right]  \sin (2 \kappa r)-2 \kappa r \cos (2 \kappa r) \Big\},
    \label{Isotropic Background Potential for scalar}
\end{equation}
and
\begin{equation}
   V^p_\text{bkg}= +\frac{G_p^2}{16\pi^3 m^2_\chi r^4} \int_0^{\infty} d\kappa \;\kappa \frac{n_+(\kappa) + n_-(\kappa)}{\sqrt{\kappa^2 + m_\nu^2}}  \Big \{ (1- 2 \kappa^2 r^2)  \sin (2 \kappa r)-2 \kappa r \cos (2 \kappa r) \Big \},
    \label{Isotropic Background Potential for pseudoscalar}
\end{equation}
for scalar and pseudoscalar interactions, respectively. We note that these isotropic background potentials are derived from the scalar and
pseudoscalar interactions in Eq.~\eqref{Effective DM Interaction}. 

\subsubsection*{Cosmic neutrino background}
After neutrino decoupling, the neutrinos still retain the relativistic Fermi-Dirac  distribution, as described in ref.~\cite{Baumann:2022mni}, given by
\begin{equation}
    n_{\pm}(\kappa, \mu,T) = \frac{1}{e^{(\kappa \mp \mu)/T}+1},
    \label{Fermi-Dirac Distribution}
\end{equation}
where $T$ is the temperature of the C$\nu$B ~\footnote{In standard cosmology, the cosmic neutrino background (C$\nu$B) decouples from the thermal bath at \( T \sim 0.8\,\mathrm{MeV} \). Due to the expansion of the Universe, its temperature drops and is today approximately \( T_0 \approx 1.7 \times 10^{-4}\,\mathrm{eV} \) (\( 1.95\,\mathrm{K} \)).}, $\kappa$ is the magnitude of the spatial momentum of the neutrino, and $\mu$ is the chemical potential. From Big Bang nucleosynthesis, the degeneracy parameter is $\mu/T \ll 1$. Therefore, we make use of this argument and calculate the leading order of the background potential, neglecting $\mu$. 

Following the authors of ref.~\cite{Ghosh_2023}, we define the dimensionless integral $I_{FD}(x,b)$:
\begin{equation}
    I_{FD}(x,b) = \int_0^{\infty} dy \frac{y}{\sqrt{y^2 +x^2}} \frac{1}{e^y + 1} \text{sin}(2by).
    \label{dimless integral FD}
\end{equation}
where 
\begin{equation}
    x \equiv m_\nu/T, \phantom{and} b \equiv rT, \phantom{and} y \equiv \kappa/T.
    \label{dimensionless parameter}
\end{equation}
\noindent Substituting the integral $I_{FD}$ into the background potentials in Eq.~\eqref{Isotropic Background Potential for scalar} and Eq.~\eqref{Isotropic Background Potential for pseudoscalar}, the $V_\text{bkg}$~\footnote{In general, the potential contains the additonal factor of $C_L(T) = \frac{1}{2} \left( 1 + \frac{p_\nu}{E_\nu + m_\nu}\right)$, accounted for chiral projection of C$\nu$B into their active component (see ref.~\cite{Ghosh_2023}). For massive Majorana neutrinos, there is no need to include the factor of chiral projection $C_L(T) = 1$.} can be expressed as
\begin{subequations}
    \begin{align}
    V_\text{bkg}^s(r) &= +\frac{G_s^2}{8\pi^3 m^2_\chi r^4}T \left[ (1 - 2b^2x^2) I_{FD} - b \frac{\partial}{\partial b} I_{FD} + \frac{b^2}{2}  \frac{\partial^2}{\partial b^2} I_{FD} \right]: \phantom{an}  \textbf{Scalar},
    \tag{\theequation a} \label{general form of FD potential of scalar}\\
     V_\text{bkg}^p(r) &= +\frac{G_p^2}{8\pi^3 m^2_\chi r^4}T \left[ I_{FD} - b \frac{\partial}{\partial b} I_{FD} + \frac{b^2}{2}  \frac{\partial^2}{\partial b^2} I_{FD} \right]: \phantom{an}  \textbf{Pseudoscalar}. 
    \tag{\theequation b} \label{general form of FD potential of pseudoscalar}
    \end{align}
\end{subequations}
The integral $I_{FD}$ cannot be computed analytically and must be evaluated numerically. Nevertheless, we can approximate it in the relativistic limit ($m_\nu \ll T$) and nonrelativistic limit ($m_\nu \gg T$). Their behaviors are shown as follows.

\subsubsection*{Relativistic limit}
In the relativistic limit ($m_\nu \ll T$), the leading term of the background potentials for both the scalar and pseudoscalar interaction in the long-range limit $(r \gg T^{-1})$ are 
\begin{equation}
    V^{s,p}_\text{bkg}(m_\nu \ll T, r \gg T^{-1}) \simeq +\frac{3G_{s,p}^2}{32 \pi^3 m_\chi^2 r^5}. \label{Massless:long range} 
\end{equation}
In the short-range limit $(r \ll T^{-1})$, the background potentials are 
\begin{align}
    V^{s,p}_\text{bkg}(m_\nu \ll T, r \ll T^{-1}) \simeq -\frac{7\pi G_{s,p}^2 }{720 m_\chi^2} \frac{T^4}{r}.\label{Massless:short range} 
\end{align}

\begin{figure}[H]
\centering
\includegraphics[width=0.8\textwidth]{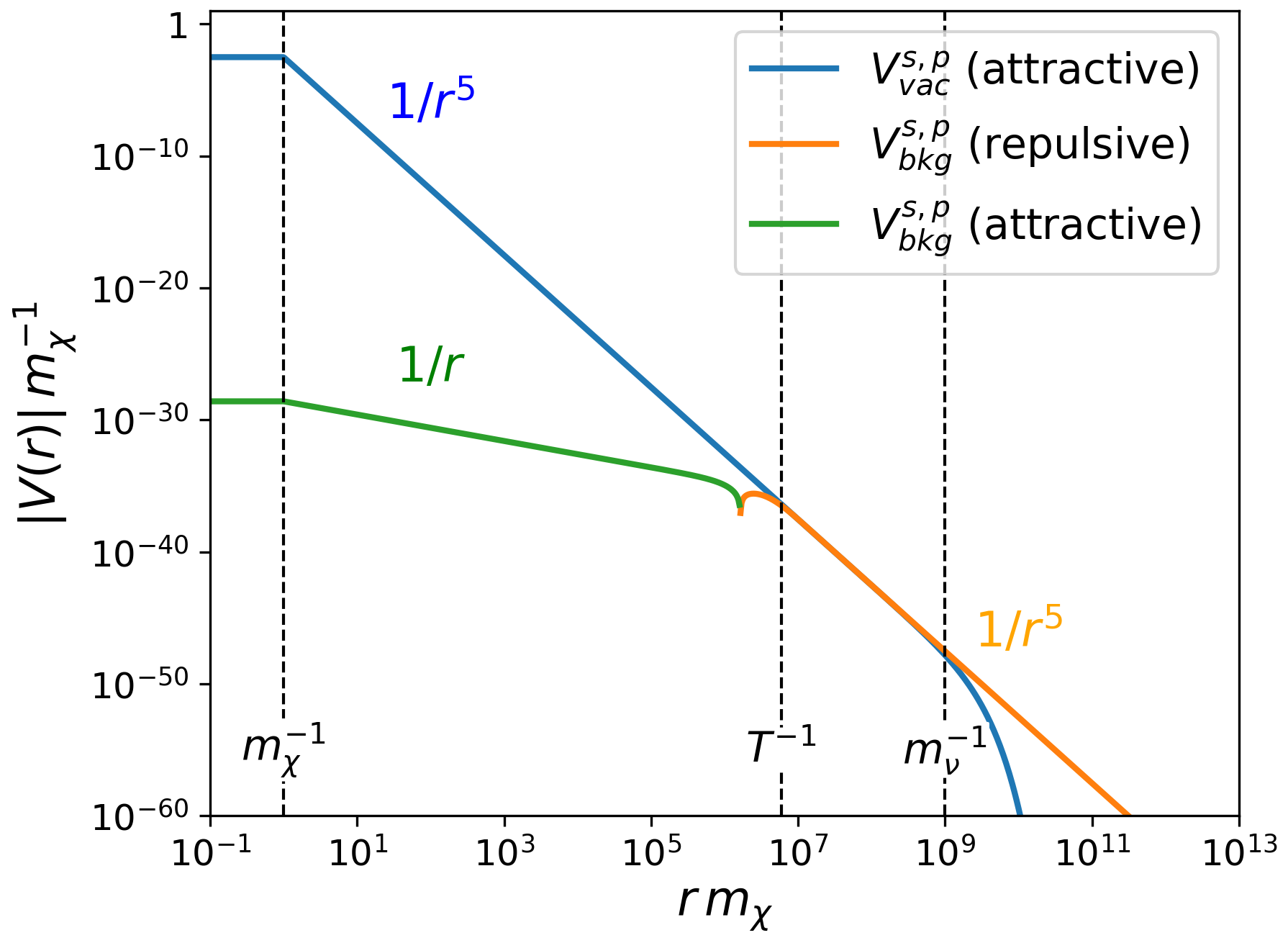}
\caption[Short-form caption]{Magnitude of the vacuum potential $V^{s,p}_\text{vac}(r)$ (blue) and background potential $V^{s,p}_\text{bkg}(r)$ (orange) from both the scalar and the pseudoscalar interactions. The potential is plotted for $m_\chi = 10^{3}$ eV, $m_\nu = 10^{-6}$ eV, and $T = 1.7 \times 10^{-4}$ eV. The potential is in the relativistic limit ($T \gg m_\nu$) and the change in behavior is seen at $rm_{\chi} \sim 10^6$. In this limit, the scalar potential is the same as the pseudoscalar.}
\label{fig:neutrino potential relativistic comparison}
\end{figure}

At the leading term, the background potential from the scalar and pseudoscalar fields exhibits the same form. It shows a repulsive $1/r^5$ profile in the long-range limit and an attractive $1/r$ profile in the short-range limit. Notably, at the long-range limit ($r \gg T^{-1}$), $V_\text{bkg}$ is repulsive, in direct contrast to the attractive nature of $V_\text{vac}$. As shown in figure~\ref{fig:neutrino potential relativistic comparison}, in the relativistic limit ($m_\nu \ll T$), the $V_\text{vac}$ is screened off by the C$\nu$B over the distance range $T^{-1} \ll r \ll m_\nu^{-1}$. At longer ranges, specifically for $r \gg m_\nu^{-1}$, the background potential $V_\text{bkg}$ dominates, overpowering the exponentially suppressed vacuum potential $V_\text{vac}$. These findings are consistent with the results derived from an effective field theory (EFT) framework in ref.~\cite{Fichet:2017bng,Costantino:2019ixl,Costantino:2020bei}.

\subsubsection*{Non-relativistic limit}
In the NR limit ($m_\nu \gg T$), the background potentials in the long-range limit $(r \gg T^{-1})$ are given as 
\begin{align}
    V^s_\text{bkg}(m_\nu \gg T, r \gg T^{-1}) &\simeq  - \frac{G^2_s}{64 \pi^3 m^2_\chi}  \frac{m_\nu}{Tr^5},  \label{NR: potential of scalar-long-range} \\ 
    V^p_\text{bkg}(m_\nu \gg T, r \gg T^{-1}) &\simeq - \frac{5G^2_p}{64 \pi^3 m^2_\chi}  \frac{1}{m_\nu T r^7} . \label{NR: potential of pseudoscalar-long-range}
\end{align}
On the other hand, the background potentials in the short-range limit $(r \ll T^{-1})$ are given as
\begin{align}
    V_\text{bkg}^s(m_\nu \gg T, r \ll T^{-1}) &\simeq  -\frac{3 \zeta(3)G^2_s}{4\pi^3m^2_\chi}  \frac{m_\nu T^3}{r} , \label{NR: potential of scalar-short-range} \\
    V^p_\text{bkg}(m_\nu \gg T, r \ll T^{-1}) &\simeq - \frac{15 \zeta(5) G^2_p}{4\pi^3 m^2_\chi} \frac{T^5}{m_\nu r}. \label{NR: potential of pseudoscalar-short-range}
\end{align}

\begin{figure}[h]
\centering
\includegraphics[width=0.8\textwidth]{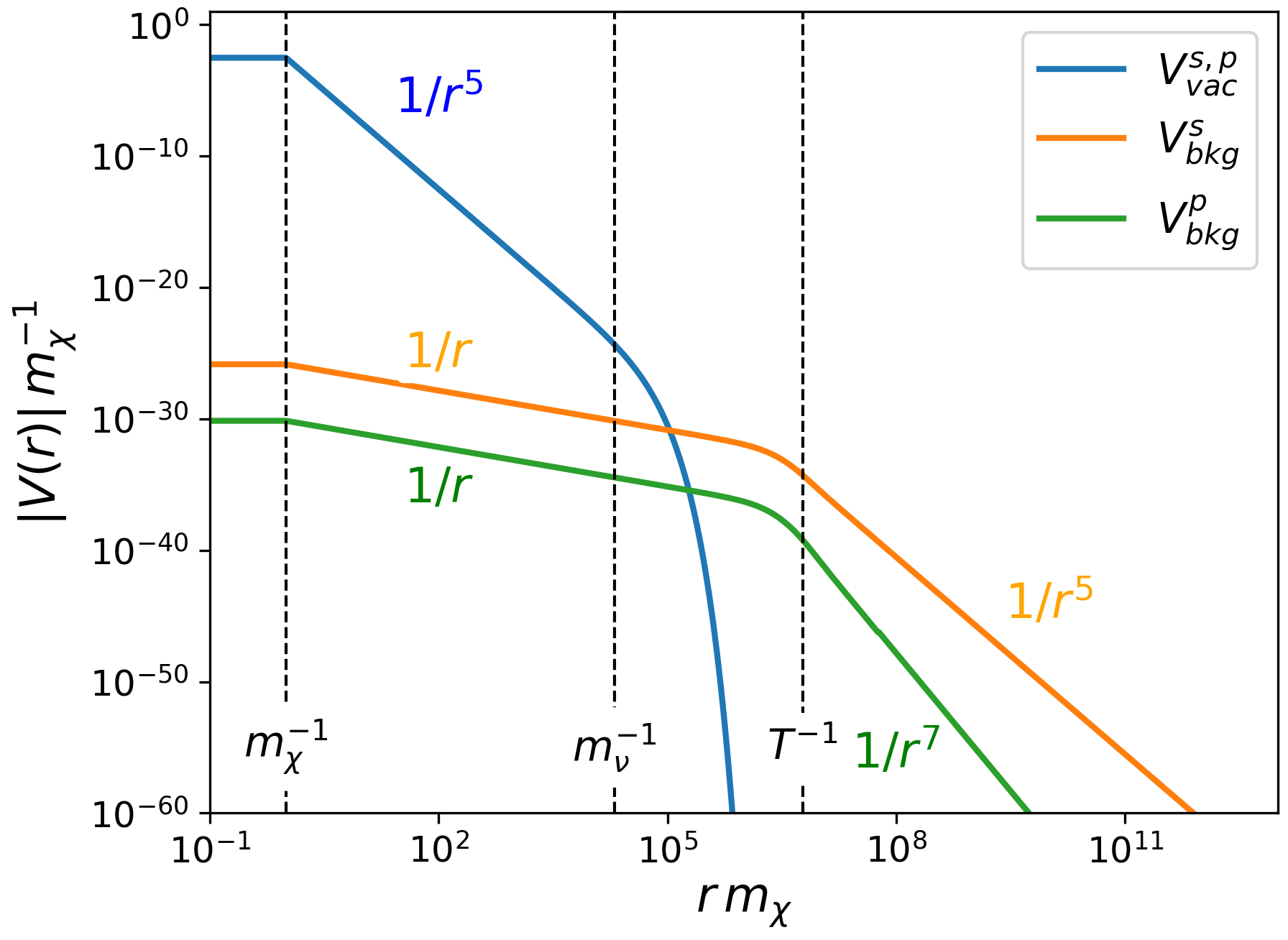}
\caption[Short-form caption]{Magnitude of the vacuum potential $V^{s,p}_\text{vac}(r)$ (blue), along with background potential $V^{s}_\text{bkg}(r)$ (orange) and $V^{p}_\text{bkg}(r)$ (green) from both the scalar and pseudoscalar interactions, respectively. The potential is plotted for $m_\chi = 10^{3}$ eV, $m_\nu = 0.05$ eV, and $T = 1.7 \times 10^{-4}$ eV. The potential is in the non-relativistic limit ($m_\nu \gg T$), showing the suppressed pseudoscalar potential compared to scalar}
\label{fig:neutrino potential NR comparison}
\end{figure}

In the non-relativistic limit ($m_\nu \gg T$), both potentials $V_\text{bkg}$ exhibit different r-dependent profiles. In the long-range limit, the $V_\text{bkg}$ from scalar interaction exhibits a $1/r^5$ profile. While the $V_\text{bkg}$ from pseudoscalar interaction exhibits a $1/r^7$ profile. In the short-range limit, both exhibit a $1/r$ profile. Therefore, the $V_\text{bkg}$ from the pseudoscalar interaction is more suppressed compared to the $V_\text{bkg}$ from the scalar interaction due to the factor $m_\nu T^{-1} \gg 1$. The NR limit ($m_\nu \gg T$) of the neutrino potential is shown in figure~\ref{fig:neutrino potential NR comparison}.

Both $V_\text{bkg}$ are also attractive, which is similar to $V_\text{vac}$ being attractive. Interestingly, the $V_\text{bkg}$ from C$\nu$B is not exponentially-suppressed at the large distance, $r \gg m_\nu^{-1}$. This is because the number density of the C$\nu$B (term in $V_\text{bkg}$) is always proportional to $T^3$. On the other hand, the vacuum potential $V_\text{vac}$ is suppressed by the factor of $e^{-2m_\nu r}$ for both the scalar and pseudoscalar interaction. Therefore, the neutrino force between two DM particles will be dominated by the $V_\text{bkg}$ for massive neutrinos at a long range ($r \gg m_\nu^{-1}$). While in the short range ($r \ll m_\nu^{-1}$), the potential is dominated by the $V_\text{vac}$ as a function of $1/r^{5}$. 

\subsection{Comparison between vacuum and background potential}
\label{subsec: implication of bkg on DM scattering}
To sum up, we have derived both the vacuum potential $V_\text{vac}$ and background potential $V_\text{bkg}$ for the scalar and pseudoscalar interaction. The general expression for $V_\text{vac}$ is given in appendix~\ref{appendix: Calculation of the massive neutrino vacuum potential}. Potentials from both interactions exhibit an attractive $1/r^5$ profile in the massless limit. The background potential $V_\text{bkg}$ is computed from the case of the cosmic neutrino background. The general $V_\text{bkg}$ expression for the scalar and pseudoscalar is given as Eq.~\eqref{general form of FD potential of scalar} and Eq.~\eqref{general form of FD potential of pseudoscalar}, respectively, which is valid for arbitrary neutrino mass and distance. The asymptotic behaviors of the potential are summarized in the table~\ref{tab:comparing force}.

\begin{table}[h]
    \centering
    \renewcommand{\arraystretch}{1.4}
    \begin{tabular}{|>{\centering\arraybackslash}p{2.8cm}|
                    >{\centering\arraybackslash}p{2.2cm}|
                    >{\centering\arraybackslash}p{2.2cm}|
                    >{\centering\arraybackslash}p{2.5cm}|
                    >{\centering\arraybackslash}p{2.5cm}|} \hline
        \multirow{2}{*}{Interaction} & 
        \multicolumn{2}{c|}{Relativistic Limit ($m_\nu \ll T$)} & 
        \multicolumn{2}{c|}{Non-Relativistic Limit ($m_\nu \gg T$)} \\
        \cline{2-5}
        & $r \ll T^{-1}$ & $r \gg T^{-1}$ & $r \ll T^{-1}$ & $r \gg T^{-1}$ \\ \hline
        Scalar & 
        $-T^4/r$ & 
        $1/r^5$ &  
        $-m_\nu T^3/r$ & 
        \makebox[2cm][c]{$-m_\nu/(T r^5)$} \\ \hline 
        Pseudoscalar & 
        $-T^4/r$ & 
        $1/r^5$ &  
        $-T^5/(m_\nu r)$ & 
        \makebox[2cm][c]{$-1/(m_\nu T r^7)$} \\ \hline
    \end{tabular}
    \caption{The asymptotic behaviors of the background potential $V_\text{bkg}(r)$ in the relativistic and non-relativistic limits with scalar and pseudoscalar interaction Eq.~\eqref{Effective DM Interaction}.}
    \label{tab:comparing force}
\end{table}

The signs of these potentials lead to a suppression or enhancement of their total magnitude at all ranges. The significance of the neutrino background effect on dark matter (DM) scattering is therefore determined by the sign and magnitudes of $V_{\text{vac}}$ and $V_{\text{bkg}}$. This is particularly important for low-energy DM scattering, since we need to understand if the screening or the enhancement is strong enough to have a noticeable impact. In general, the potential becomes significant for DM scattering when its magnitude is comparable to or dominates the particle's kinetic energy, which occurs at the short range where the potential is largest. We discuss the impact of the cosmic neutrino background in each of the following limits below. 

In the non-relativistic limit ($m_\nu \gg T$), the $V_{\text{vac}}$ is in attractive $1/r^5$ ($r \ll m_\nu^{-1}$) form, whereas $V_{\text{bkg}}$ is in the attractive form of $1/r$ ($r \ll T^{-1}$). Therefore, the potential $V_{\text{vac}}$ dominates $V_{\text{bkg}}$ in the short range ($r\ll T^{-1}$). Following from figure~\ref{fig:neutrino potential NR comparison}, this implies that the cosmic neutrino background effect can be negligible in this limit.

\begin{figure}[h]
\centering
\includegraphics[width=0.8\textwidth]{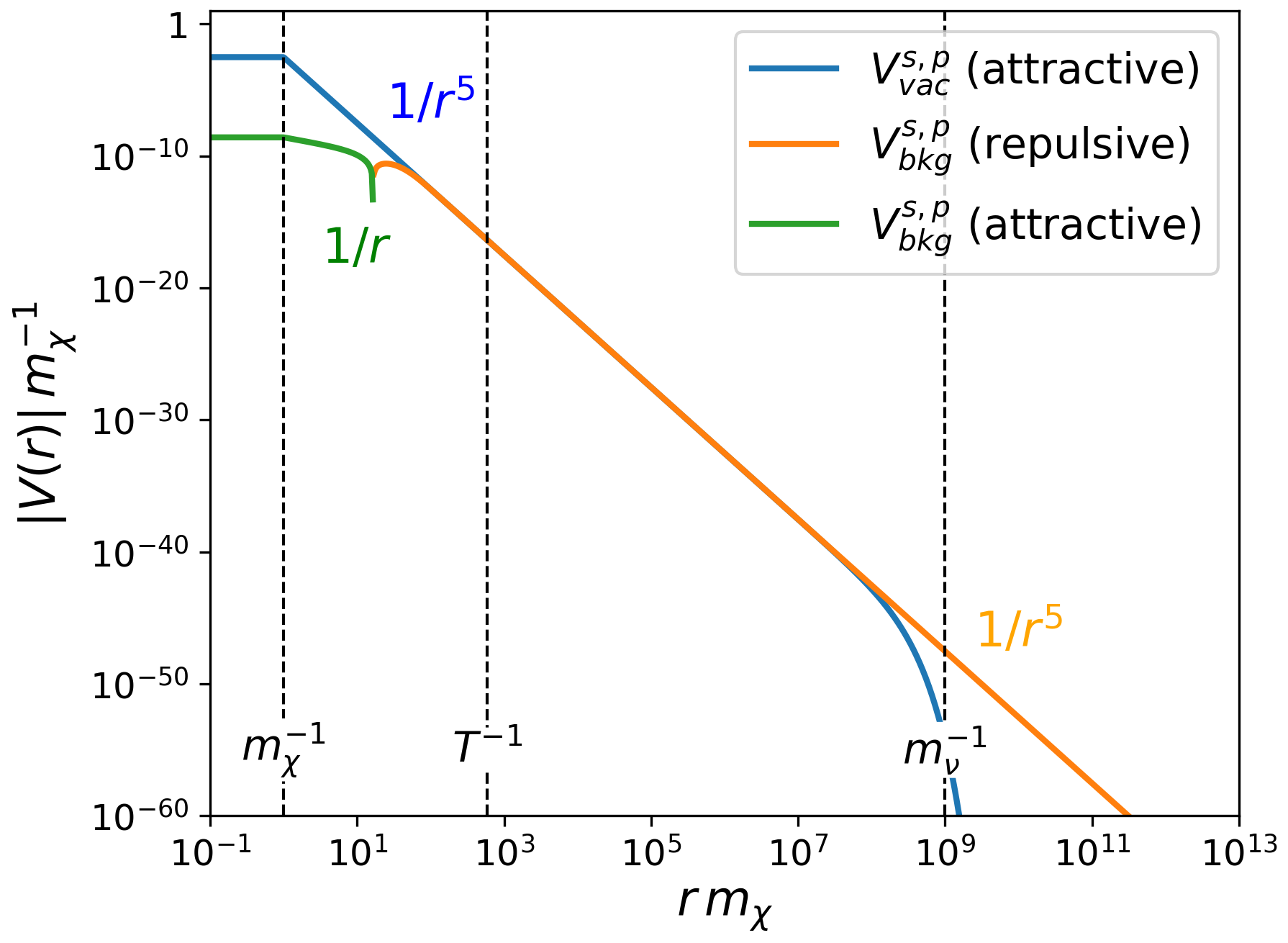}
\caption[Short-form caption]{Magnitude of the vacuum potential $V^{s,p}_{\text{vac}}(r)$ (blue) and background potential $V^{s,p}_{\text{bkg}}(r)$ (orange) from both the scalar and pseudoscalar interactions. The potential is plotted for $m_\chi = 10^{3}$ eV, $m_\nu = 10^{-6}$ eV, and $T = 10^4 \times T_0$. The potential is in the relativistic limit ($T \gg m_\nu$). In this limit, $V_{\text{vac}}$ is completely opposed to $V_{\text{bkg}}$ up to the short range.}\label{fig:neutrino potential comparison high T}
\end{figure}

In the relativistic limit ($m_\nu \ll T$), the $V_{\text{vac}}$ and $V_{\text{bkg}}$ exhibits completely the same form of $1/r^5$ at $T^{-1} \ll r \ll m_\nu^{-1}$, but with opposite sign. The background potential $V_{\text{bkg}}$ changes its form to $1/r$ at the scale $T^{-1}$ toward $m_\chi^{-1}$. For the present time \text{C$\nu$B} ($T_0 \approx 1.7 \times 10^{-4}$ eV), the potential $V_{\text{vac}}$ dominates $V_{\text{bkg}}$ in the short distance. However, since the DM scattering appears as well in the earlier time, we should also consider the temperature of the cosmic neutrino background at each cosmic time. This case is illustrated in figure~\ref{fig:neutrino potential comparison high T} where $T = 10^4 \times T_0$. Quantitatively, as temperature $T$ increases, the scale $T^{-1}$ approaches the scale $m_\chi^{-1}$. This extends the range of the $1/r^5$ form of  $V_{\text{bkg}}$ to regimes where it becomes comparable to $V_{\text{vac}}$, leading to a complete screening of the potential for dark matter with mass $m_\chi \lesssim T$. 

In conclusion, the effect of the cosmic neutrino background becomes significant when the DM mass is low enough (comparable to the scale $T$ or $m_\nu$) such that $V_{\text{bkg}}$ is comparable to $V_{\text{vac}}$. For large DM masses, the scattering behavior resembles the vacuum case, with the background potential playing a negligible role. Hence, the significance of the neutrino background in DM scattering is fundamentally determined by the mass hierarchy of $m_\chi$, $m_\nu$, and $T$.

\section{Phenomenology of the self-interacting dark matter}
\label{sec: phenomenology}
In this section, we study the effect of the C$\nu$B on the phenomenology of the proposed DM-neutrino interaction. In particular, the interplay between vacuum and background potential modifies the DM self-scattering and annihilation. First, we review the computational method for DM self-scattering and annihilation in section~\ref{subsec: self-scattering} to establish the practical framework in low-energy DM scattering. Moreover, we study the parameter space, addressing the core-cusp problem, in section~\ref{subsec: Parameter space} with relevant parameters: interactions, neutrino mass, and the C$\nu$B temperature. Next, we examine the Sommerfeld enhancement in section~\ref{sec:Effect of Cosmic Neutrino Background on Sommerfeld enhancements}. Lastly, we discuss the possible UV model. This also includes examining the constraints from the SM decay channels through possible UV interaction.

\subsection{DM self-scattering and annihilation via neutrino force}
\label{subsec: self-scattering}
In this section, we review the computational method for determining cross-sections in DM self-scattering and the Sommerfeld Enhancement factor in DM annihilation, resulting from the presence of the neutrino force. Given that the neutrino potential originates from the effective dimension-5 operator in Eq.~\eqref{Effective DM Interaction}, the potential is singular and not well-behaved. We also address the challenges posed by such singular potentials and outline appropriate treatments.  

\subsubsection*{Computational method for DM-neutrino interaction}

\begin{figure}[h]
\centering
\includegraphics[width=0.45\textwidth]{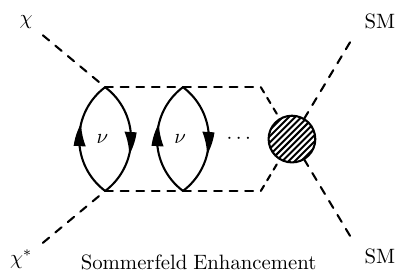}
\hfill
\includegraphics[width=0.45\textwidth]{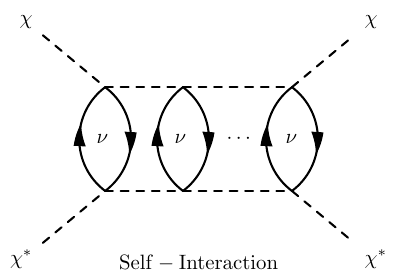}
\caption[Short-form caption]{Multiple exchange of the neutrino/antineutrino pair, resulting in the ladder diagram.}
\label{fig:Multiple exchange of the neutrino pair}
\end{figure}

We consider interactions involving slowly moving DM particles exchanging a pair of neutrinos/antineutrinos through $\chi \chi^* \rightarrow \chi\chi^*$. The general form of the potential is given by Eq.~\eqref{general form of FD potential of scalar} and~\eqref{general form of FD potential of pseudoscalar} for the scalar and pseudoscalar scenarios, respectively. In a non-relativistic scattering, multiple neutrino pairs may be exchanged while undergoing the DM annihilation or self-scattering -- depicted by ladder diagrams in figure~\ref{fig:Multiple exchange of the neutrino pair}.

Such interactions necessitate resummation beyond Born's approximation if scattering occurs outside the perturbative regime. Previous analyses, such as ref.~\cite{Orlofsky_2021}, have demonstrated substantial effects from multiple exchanges, particularly for repulsive neutrino potentials scaling as $1/r^5$.
Therefore, ladder diagrams at all loop levels must be summed to account accurately for non-perturbative effects. In QFT, these non-perturbative effects are treated by solving the Bethe-Salpeter equation (ref.~\cite{Iengo_2009}). However, considering the low-energy and non-relativistic nature of DM particles, these effects can be effectively captured using non-relativistic QM. Specifically, the non-perturbative interactions are accounted for by solving the Schrödinger equation in the reduced two-particle system (See ref.~\cite{Petraki_2015} for more detail).
The potential between DM particles alters the wave function, subsequently affecting self-scattering and annihilation cross sections.

The Schr\"{o}dinger equation for the two-particle reduced DM system is given by
\begin{equation}
    \frac{1}{r^2} \frac{d}{dr} \left( r^2 \frac{dR_{l}}{dr} \right) + \left( k^2 - \frac{l(l+1)}{r^2} - 2\mu V(r) \right) R_{l} = 0,
    \label{SE: DM two-particle system}
\end{equation}
where $R_{l}$ denotes the radial wave function with angular momentum $l$, $\mu = m_\chi/2$ represents reduced mass, and $k$ is the relative momentum. This can be derived from expanding the wave function in the form $\psi(r,\theta) = \sum_{l} R_{l}(r) P_l(\cos \theta)$, with $P_l$ being Legendre polynomials. DM scattering is quantified through the differential cross section by $d\sigma = \abs{f(\theta)}^2 d \Omega,$ where the scattering amplitude $f(\theta)$ is decomposed into partial wave modes $l$ and phase shifts $\delta_l$ as
\begin{equation}
    f(\theta) = \frac{1}{2ik} \sum_{l = 0}^{\infty} (2l + 1)(e^{2i\delta_l} - 1) P_l(\cos \theta).
\end{equation}
The cross-section is then given by
\begin{equation}
     \sigma = \frac{4 \pi}{k^2} \sum_{l = 0}^{\infty} (2l + 1) \sin^2\delta_l.
     \label{cross section for l}
\end{equation}
To compute the phase shift $\delta_l$, we start by considering a simple method of solving the radial Schr\"{o}dinger Eq.~\eqref{SE: DM two-particle system}. The boundary condition demands that $u_{l}(r) \equiv r R_{l}$ must vanish at $r = 0$. Near the origin, the wave function behaves as $R_l \propto r^l$. At a large distance, where the potential vanishes, the asymptotic behavior of $R_{l}$ becomes 
\begin{equation}
  R_{l}(r) \propto   \cos (\delta_l) j_l(kr) - \sin (\delta_l) n_l(kr),
\end{equation}
where $j_l$ and $n_l$ are the spherical Bessel functions of the first and second kinds, respectively. Following the approach outlined in ref.~\cite{Chu_2020}, we define
\begin{equation}
    t_{l}(r) = \frac{j_l(kr) \left( \frac{R'_{l}(r)}{R_l(r)} - \frac{l}{r} \right) + k j_{l+1}(kr)}{n_l(kr) \left( \frac{R'_{l}(r)}{R_l(r)} - \frac{l}{r} \right) + k n_{l+1}(kr) }.
\end{equation}
This simplifies Eq.~\eqref{SE: DM two-particle system} into a first-order differential equation
\begin{equation}
    \frac{dt_l}{dr} = -km_\chi r^2 V(r) \left( j_l(kr) - t_l n_l(kr)\right)^2
    \label{cross-section ODE}.
\end{equation}
Since $R_l \propto r^l$ and vanishes at $r \rightarrow \infty$, we have the boundary condition of $t_l(r = 0) = 0$ and $t_l(r \rightarrow \infty) \rightarrow \tan (\delta_l)$. 
This provides a straightforward method to determine the phase shift and cross-section by solving Eq.~\eqref{cross-section ODE} with the given boundary conditions. 

DM annihilation predominantly occurs near the origin at $r = 0$. The presence of the long-range force distorts the wave function at $r = 0$, modifying the annihilation cross-section. The Sommerfeld enhancement factor $S_l$ quantifies this modification and is defined as the ratio of the annihilation cross-section distorted by the neutrino force and that in the absence of the neutrino force. As demonstrated in ref.~\cite{Arkani_Hamed_2009}, higher partial waves ($l \ge 1$) do not significantly contribute to the factor $S_l$ if the potential remains less singular than $1/r$ near the origin. Therefore, we restrict our analysis to the $l = 0$ mode. The Sommerfeld enhancement factor is given by 
\begin{equation}
    S = \abs{\frac{\psi_{\text{distorted}}(r = 0)}{\psi_{\text{free}}(r = 0)}}^2 =\lim_{r \rightarrow 0} \abs{\frac{R(r)}{R_{\text{free}}(r)}}^2 = \abs{u'(0)}^2,
    \label{Sommerfeld Enhancement factor}
\end{equation}
where $R(r)$ and $R_{\text{free}}(r)$ denote the radial part of the wavefunction of Eq.~\eqref{SE: DM two-particle system} with and without neutrino force, respectively. 

Following ref.~\cite{Arkani_Hamed_2009}, the amplitude of $u(r)$ at large distances ($r \rightarrow \infty$) must match the free solution $u_{\text{free}}(r) \propto \sin(kr)/k$. Thus, Eq.~\eqref{SE: DM two-particle system} is solved with the initial conditions $u(0)=0$ and $u'(0)= c$, adjusting the constant $c$ to match the amplitude at infinity. Since Eq.~\eqref{SE: DM two-particle system} is linear in $u$, we conveniently set $u'(0)=1$, solve for the amplitude $A_u$, and use it to compute the Sommerfeld enhancement factor as
\begin{equation}
    S = \abs{\frac{1}{k A_u}}^2.
    \label{Sommerfeld Enhancement factor computed}
\end{equation}

\subsubsection*{Renormalized neutrino potential} 
Singular potentials $V \propto 1/r^n$ with $n > 2$ invalidate the Schrödinger equation, as they lead to unbounded Hamiltonians and infinitely deep bound states.
This causes particles falling to an infinitely small $r$ due to the attractive short-distance potential. Such a singular behavior also emerges in neutrino potentials, where both $V_\text{vac} \sim 1/r^5$ and $V_\text{bkg} \sim 1/r^5$ become singular in the limit $T \gg m_\chi$. This issue can be resolved by treating the quantum mechanical model of scattering as a low-energy effective theory derived from the underlying UV theory (see ref.~\cite{Bedaque_2009}). 
Below a certain scale $R$, this non-relativistic description and short-distance singular potential are no longer valid. However, the Schr\"{o}dinger equation remains applicable provided the appropriate renormalization procedure, which encodes the UV physics, is performed.

In our study, we renormalize the Schr\"{o}dinger equation by adopting the Wilsonian renormalization group~\cite{lepage1997renormalizeschrodingerequation}. Previous literature has employed a similar approach (See refs. ~\cite{Coy_2022, Ferrante:2025lbs, Bellazzini_2013}), where the singular potential is regulated by a cutoff $R$ accompanied by a series of local operators representing UV physics. Since these local terms encode UV physics, the details of the counterterm are irrelevant. We adjust it as a flat potential with height $V_0$.

Following this method, the potential is regulated using the parameter $V_0$ and the cutoff $R$. The physical results thus depend on the arbitrariness of a single parameter for a central $1/r^n$ singular potential~\cite{Beane_2001}. We choose the potential at the radius $R$ to be $V_0$, while also imposing continuity with the non-singular long-range part. Therefore, the potential is given by
\begin{equation}
    V(r) = 
    \begin{cases}
        V_{\rm vac}(r) + V_{\rm bkg}(r), & \text{for } r > R \\
        V_0 = V_{\rm vac}(R) + V_{\rm bkg}(R),  & \text{for } r \le R.
    \end{cases} 
\end{equation}
Note that physical observables depend on a single parameter $R$, corresponding to the unknown UV physics. In other words, determining $R$ involves matching the effective theory to a known UV completion model. As a result, the universal character of the cross-section per DM mass is no longer presented. In this work, we set $R = m_\chi^{-1}$, the scale beyond which NR scattering becomes invalid \footnote{ref.~\cite{Ferrante:2025lbs} employs a DM relative velocity-dependent cutoff $R^{-1} \sim m_\chi v \ll G^{-1}$ which is applicable to NR scattering. However, in our work, we simply place the cutoff at $R^{-1} \sim m_\chi \ll G^{-1}$.}.

\subsection{Parameter space addressing the core-cusp problem} 
\label{subsec: Parameter space}
We now derive the parameter space for self-interacting DM mediated by the neutrino force, incorporating effects from the cosmic neutrino background. Through this analysis, we consider the DM mass to be larger than the neutrino mass $m_\chi > m_\nu$ such that the neutrino mass remains the relevant degree of freedom. The effect of neutrino mass and C$\nu$B's temperature is investigated in each parameter set. The self-scattering cross-section is obtained from solving the Schr\"{o}dinger equation through Eq.~\eqref{cross section for l} and Eq.~\eqref{cross-section ODE}. We highlight the region where the cross-section per DM mass lies within $0.03\;\text{cm}^2/\text{g} \le \sigma_{\chi\chi^* \rightarrow \chi \chi^*} / m_\chi \le 1 \; \text{cm}^2/\text{g}$ as this range may be relevant for addressing various small-scale structure challenges~\cite{Kaplinghat_2016}. The corresponding parameter spaces are shown in figure~\ref{fig: Parameter space for scalar/pseudoscalar interaction in the relativistic limit} – \ref{fig: Parameter space for $V_{vac} + V_{bkg}$ of the scalar interaction, varying temperature} and~\ref{fig: full theory s-channel parameter space}.

For simplicity, we consider the contribution from a single neutrino mass. Results are presented for both the limit of massless ($m_\nu = 0$) and the massive neutrino ($m_\nu \simeq 0.05 \;\text{eV})$. The latter represents a lower bound for the heaviest neutrino mass derived from neutrino oscillation experiments,  \(\Delta m_{31}^2 \simeq 2.5 \times 10^{-3}\,\mathrm{eV}^2\)~\cite{ParticleDataGroup:2024cfk} where we take the lightest neutrino mass to be \(m_1 \ge 0\). The C$\nu$B's temperature in our work ranges from the present-day value $T =T_0 \simeq 1.7 \times 10^{-4}$ eV back to earlier epochs, allowing us to examine both relativistic and non-relativistic limits of the potentials. The general results are summarized below.

A general trend toward larger cross-section per DM mass for larger coupling $G$ is observed in every sample parameter set, due to stronger potentials. The resulting cross-section goes as $G^4$, confirming that the Born approximation is valid for the DM scattering via the neutrino force for our interaction in Eq.~\eqref{Effective DM Interaction}. Note that this contrasts the $G^{4/3}$ dependence in ref.~\cite{Orlofsky_2021}, where cross-sections are analytically derived from repulsive neutrino potentials of the form $1/r^5$. Such an effect is referred to arise from the resummation of multiple neutrino exchange. The differing signs of the potential explain these discrepancies. Our study employs an attractive $1/r^5$  potential, regulated via the flat potential.

In the NR limit, the neutrino background potential $V_{\text{bkg}}$ has minimal impact for both scalar and pseudoscalar interactions, as the impact on the cross section is dominated by the vacuum potential $V_{\text{vac}}$ at short range ($r \ll T^{-1}$). In the relativistic limit, background potentials $V_{\text{bkg}}$ for both interactions share the same forms, making their parameter spaces indistinguishable. In a regime where DM mass is smaller than the temperature, $m_\chi \lesssim T$, the background potential $V_{\text{bkg}}$ cancels the vacuum potential $V_{\text{vac}}$ at short range ($r \ll T^{-1}$), and effectively reduces the scattering cross-section.

\subsubsection*{Scalar and pseudoscalar interactions}
\begin{figure}[h]
    \centering    \includegraphics[width=0.7\linewidth]{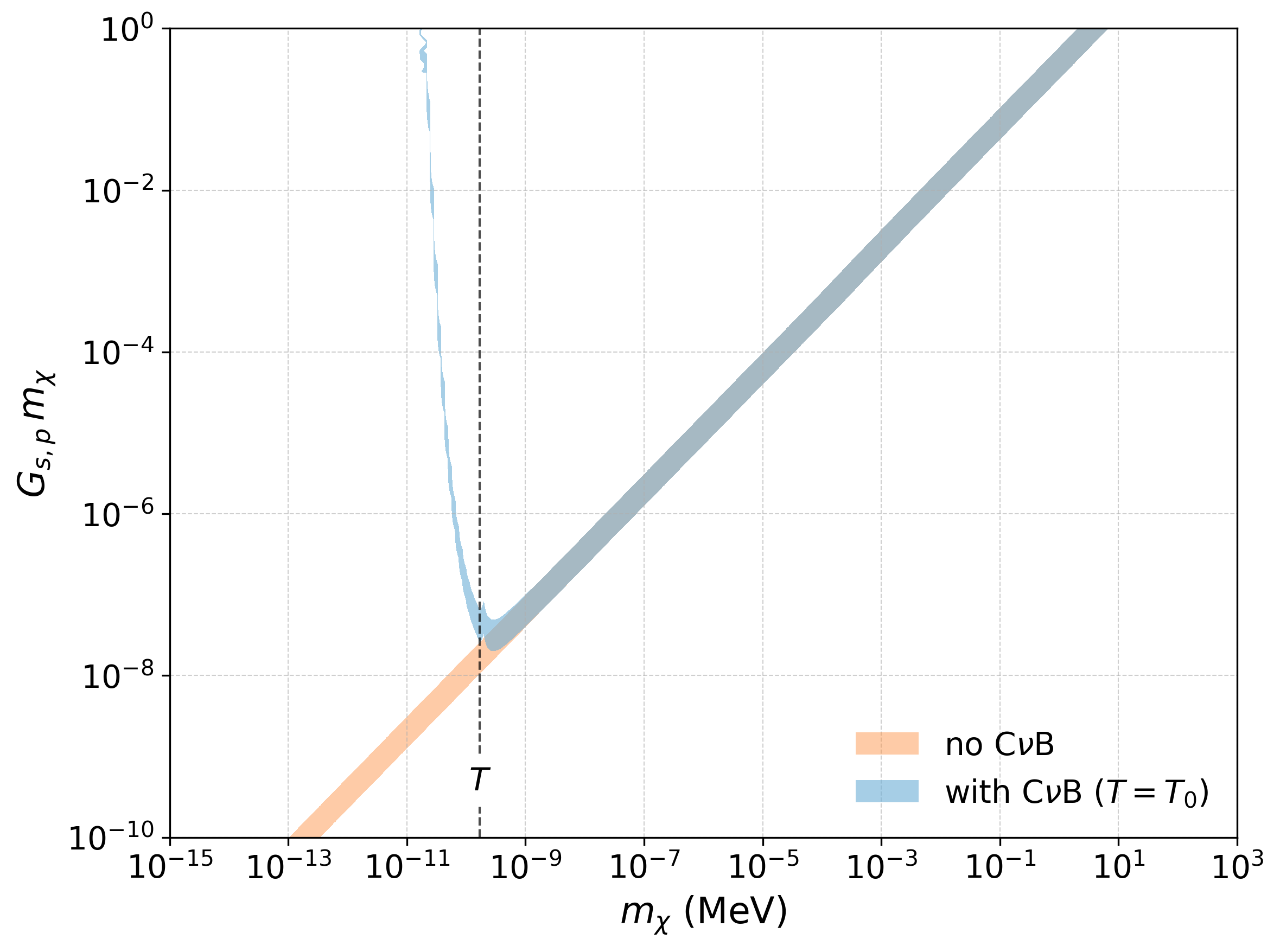}
\caption[Short-form caption]{The impact of the cosmic neutrino background on the parameter space which can address the core-cusp problem in the relativistic case ($m_\nu = 0$ $\text{eV}$, $T = T_0$). The orange-shaded band corresponds to the sufficient dark matter self-interaction via neutrinos with no background effect. The blue-shaded band incorporates the cosmic neutrino background effect. The blue-shaded band is shifted up due to the screening between $V_{\rm vac}$ and $V_{\rm bkg}$.}
\label{fig: Parameter space for scalar/pseudoscalar interaction in the relativistic limit}
\end{figure}

\begin{figure}[h]
\centering
\includegraphics[width=0.7\linewidth]{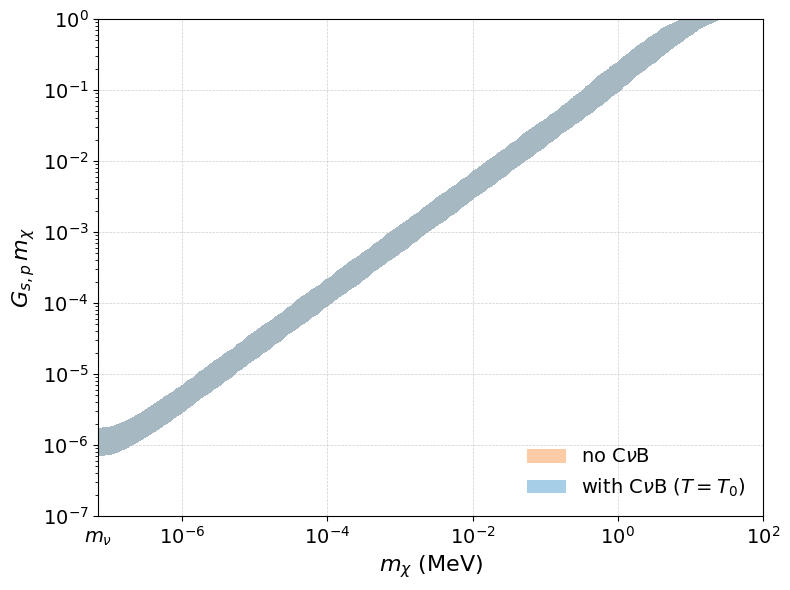}
\caption[Short-form caption]{The impact of the cosmic neutrino background on the parameter space which can address the core-cusp problem  in the non-relativistic case (orange: no C$\nu$B effect; blue: with C$\nu$B effect). In making this plot, we take $m_\nu = 0.05$ eV and $T = T_0$. Note that the preferred regions of parameter space in both cases are overlapping.}
\label{fig: Parameter space for scalar/pseudoscalar interaction in the non-relativistic limit}
\end{figure}

The primary distinction between scalar and pseudoscalar interactions lies in their background potentials in the NR limit. To investigate the interaction significance, we assume scenarios where either one interactions is suppressed. Parameter spaces are derived for DM self-scattering under the neutrino potential for the scalar and pseudoscalar interactions. These parameter spaces, compared for scenarios with and without the C$\nu$B's contribution, are shown in figure~\ref{fig: Parameter space for scalar/pseudoscalar interaction in the relativistic limit} and figure~\ref{fig: Parameter space for scalar/pseudoscalar interaction in the non-relativistic limit} for both relativistic and  NR limits, respectively.

Since the $V_{\text{bkg}}$ takes the same form for both interactions in the relativistic limit and is insignificant for the DM scattering in the NR limit, the viable parameter spaces for both interactions coincide. Therefore, only the scalar interaction is discussed further.


\subsubsection*{Neutrino mass}
\begin{figure}[htbp]
\centering
\includegraphics[width=0.7\linewidth]{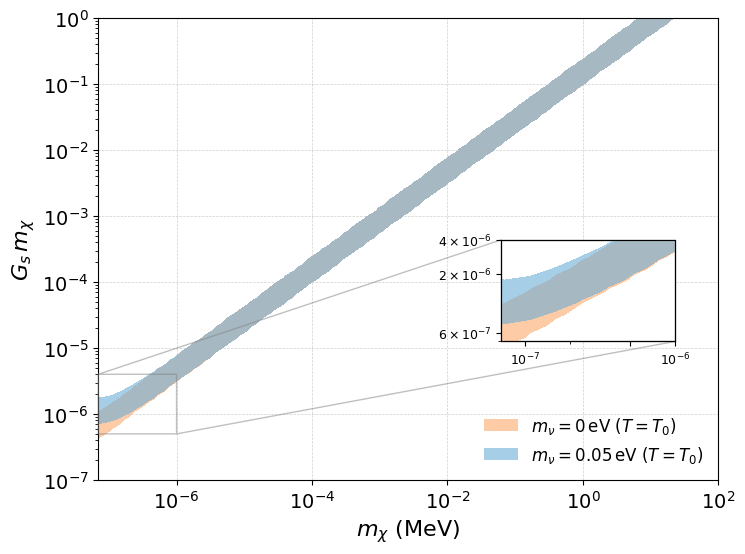}
\caption[Short-form caption]{Effect of neutrino mass on the parameter space which can address the core-cusp problem. Orange and blue shaded bands correspond to a massless neutrino ($m_\nu = 0$) and a massive neutrino ($m_\nu = 0.05$ eV) in the presence of the C$\nu$B ($T = T_0$), respectively. Around $m_\chi \sim m_\nu$, the orange band shifts upward because $V_{\rm vac}$ is exponentially suppressed. Note that for a higher temperature $T>T_0$, there is also a similar shift at $m_\chi \sim m_\nu$.
}
\label{fig: neutrino mass effect on parameter space}
\end{figure}

Neutrino mass influences DM scattering primarily through two mechanisms. First, it changes the magnitude of the background potential, as shown by the difference between relativistic and non-relativistic contributions (Table.~\ref{tab:comparing force}). Second, it induces an exponential falloff of the vacuum potential at distances $r \gg m_\nu^{-1}$. We compare the allowed parameter space, incorporating the C$\nu$B, in the massless limit $(m_\nu = 0)$ with the massive neutrino $(m_\nu = 0.05 \; \text{eV})$ case in figure~\ref{fig: neutrino mass effect on parameter space}. 

In the region with higher DM mass ($m_\chi \gg m_\nu$), the allowed regions are essentially indistinguishable. In contrast, when $m_\chi \sim m_\nu$, the allowed region shifts upwards (towards larger $G_s$) to satisfy the core-cusp constraints. This means that the potential magnitude is significantly reduced. In particular, at this DM mass range, the vacuum potential \text{$V_{\rm vac}$} exponentially decays such that the repulsive \text{$V_{\rm bkg}$} is comparable to the attractive \text{$V_{\rm vac}$}. This leads to partial screening of the net attraction and thus a ``bump'' toward larger effective coupling.


\subsubsection*{Cosmic neutrino background temperature}
As noted above, DM scattering can occur at earlier times, when the cosmic neutrino background temperature is higher. In this section, we derive parameter spaces for different ranges of temperature $T$. The result is shown in figure~\ref{fig: Parameter space for $V_{vac} + V_{bkg}$ of the scalar interaction, varying temperature} for $T = 10^4 \times T_0$ (left) and $T = 10^6 \times T_0$ (right), all within the range after the neutrino decoupling.
\begin{figure}[h]
\centering
\begin{subfigure}{0.49\textwidth}
    \centering    \includegraphics[width=\linewidth]{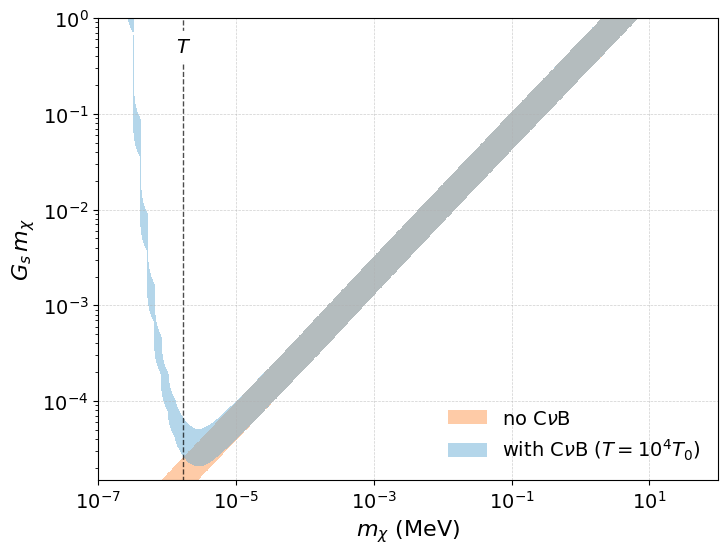}
\end{subfigure}
\hfill
\begin{subfigure}{0.49\textwidth}
    \centering   \includegraphics[width=\linewidth]{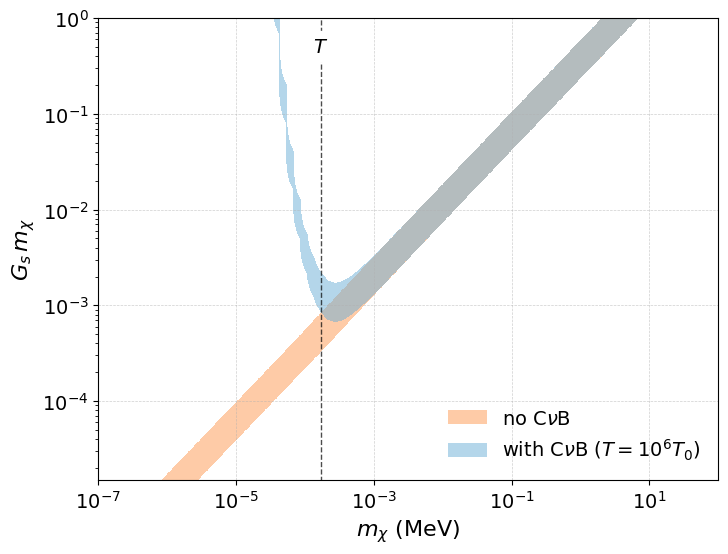}
\end{subfigure}
\caption[Short-form caption]{Impact of the cosmic neutrino background temperature ($T$) on the parameter space which can address the core-cusp problem (orange: no C$\nu$B effect; blue: with C$\nu$B effect). For higher $T$,
the C$\nu$B becomes significant at heavier $m_\chi$.
}
\label{fig: Parameter space for $V_{vac} + V_{bkg}$ of the scalar interaction, varying temperature}
\end{figure}

For higher temperatures $T$, the cosmic neutrino background affects a broader mass range, roughly $m_\chi \lesssim T$. This follows from the fact that the background potential completely screens the vacuum potential; both are in the same form of $1/r^5$ in the range $r \gg m_\chi^{-1} \gtrsim T^{-1}$, which is discussed in section~\ref{subsec: implication of bkg on DM scattering}. Thus, even relatively heavy DM can experience significant cosmic neutrino background effects at earlier times.

Besides the DM self-scattering, since the DM annihilation also happens in the presence of a neutrino background, the C$\nu$B can affect the DM annihilation process through modification in multiple neutrino exchange. This will be discussed in the next section.

\subsection{Effect of cosmic neutrino background on sommerfeld enhancements}
\label{sec:Effect of Cosmic Neutrino Background on Sommerfeld enhancements}
\begin{figure}[h]
    \centering    
    \includegraphics[width=0.7\textwidth]{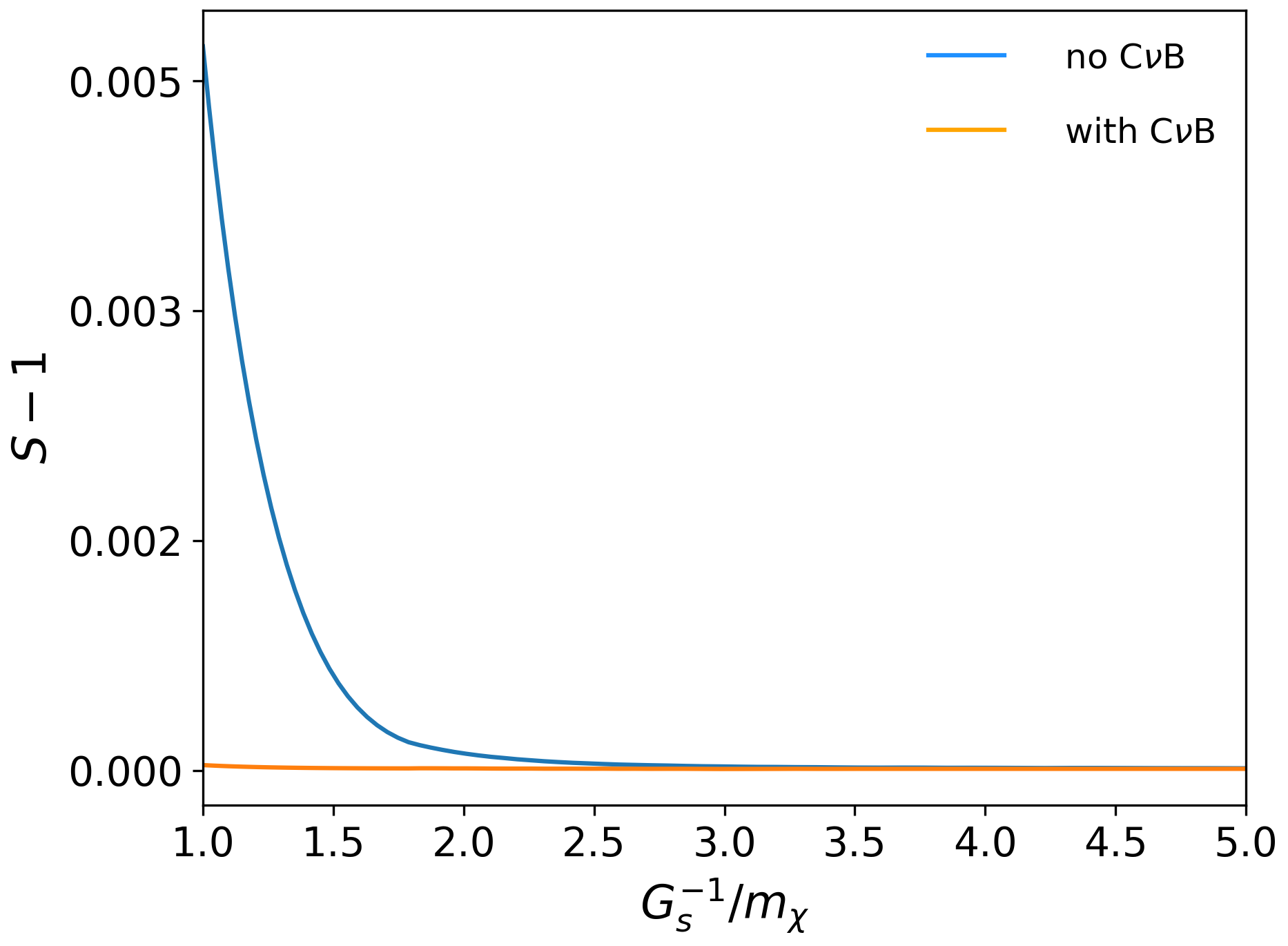}
\caption[Short-form caption]{Sommerfeld enhancement due to neutrino forces from contact interaction with $G_s < m_\chi^{-1}$. The plots are shown for $m_\chi = 10^{-2}$ MeV, $ T = 10^6 \times T_0$. The Sommerfeld enhancement is negligible for the DM annihilation. }
\label{fig: sommerfeld enhancement factor}
\end{figure}

In this section, we compute the Sommerfeld enhancement factor $S$ from the DM-neutrino effective operator in Eq.~\eqref{Effective DM Interaction}, incorporating the cosmic neutrino background effect. The factor is evaluated using Eq.~\eqref{Sommerfeld Enhancement factor computed} under the neutrino potential \text{$V = V_{\rm vac} + V_{\rm bkg}$}. 
Since the neutrino potential is singular as $1/r^5$, we regulate it as described in the section~\ref{subsec: self-scattering}. We present the factor $S-1$ with and without the background correction in figure~\ref{fig: sommerfeld enhancement factor}. The effective coupling is reliable for non-relativistic scattering ($G_s \ll m_\chi^{-1}$). Extending beyond this range would make the results sensitive to UV physics underlying the contact interaction (See ref.~\cite{Coy_2022} for possible UV channels). Therefore, we restrict the range of our plots accordingly. 

In the regime with $G_s < m_\chi^{-1}$, the factor $S$ increases with $G_S$ since a larger coupling leads to a more substantial neutrino potential and a stronger Sommerfeld enhancement effect. Without thermal correction, the factor $S-1$ is limited to be $0.005$. Including C$\nu$B correction, the factor $S-1$ is significantly reduced due to the screening of the neutrino potential by the repulsive background potential. In the non-relativistic limit ($m_\nu \gg T$), the C$\nu$B correction to the ladder diagram in the annihilation process is insignificant. While in the relativistic limit ($m_\nu \ll T$), the C$\nu$B effects can be important if the cosmic neutrino background has a higher temperature $T$. For the parameters shown above,  the enhancements completely vanish for $R = m_\chi^{-1}$. Thus, the thermal correction from the cosmic neutrino background is significant for the Sommerfeld enhancement in the annihilation. 

Note that as the potential is regulated by the cutoff $R$, the Sommerfeld enhancement is non-universal.
At $G_s > m_\chi^{-1}$, the $1/r^5$ potential ceases to be a valid non-relativistic description.
As $r \sim m_\chi^{-1}$, the momentum transfer would be comparable to $G_s^{-1}$. Within this regime, the Sommerfeld enhancement depends on the UV channel of this contact interaction. The possible UV interaction will be discussed in the next section. 

\subsection{Possible UV channel and the nature of scalar DM}
\label{subsec: UV channel}

In this section, we discuss a possible UV channel that generates the effective DM-neutrino self-interaction in Eq.~\eqref{Effective DM Interaction}, focusing on scalar DM $\chi$ coupled to active neutrinos $\nu$ such that the quadratic effective operator is generated. This effective vertex can be generated by the t-channel and s-channel mediators, similar in structure to the fermionic DM case (see ref.\cite{Xu_2022}).

\begin{figure}[htbp]
\centering
\includegraphics[width=0.7\textwidth]{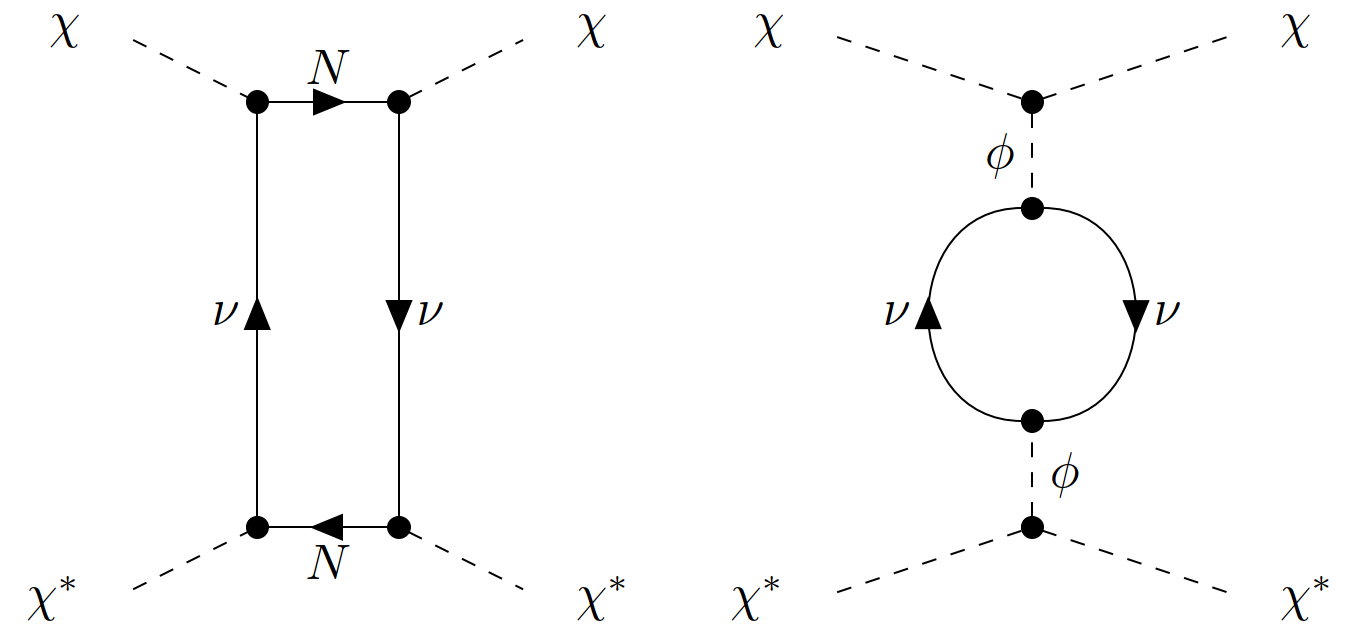}
\caption[Short-form caption]{The effective DM-neutrino interaction in Eq.~\eqref{Effective DM Interaction} is generated by the s-channel (left) and the t-channel (right) mediator via integrating out heavy fermion $N$ and heavy scalar $\phi$, respectively.}
\label{fig:Scalar DM Self-Interaction in t-channel}
\end{figure}

\newpage

For the $s$-channel mediator, the quadratic effective operator can be generated via the Yukawa interaction between a neutrino and another heavy fermion $N$:
\begin{equation}
    \mathcal{L} \supset - \chi \bar{\nu} (y_s + y_p i\gamma^5) N - M_N \bar{N} N.
    \label{s-channel UV}
\end{equation}
Integrating out the heavy fermion ($M_N \gg m_\chi$) yields the effective coupling $G_s \simeq (y^2_s + y^2_p)/M_N$ and $G_p \simeq y_sy_p / M_N$. For the $t$-channel mediator, the quadratic interaction in Eq.~\eqref{Effective DM Interaction} is generated via a scalar mediator $\phi$ with mass $M_\phi$, which couples to scalar DM $\chi$. This is given by
\begin{equation}
    \mathcal{L} \supset  - \phi \bar{\nu} (y_s + y_p i\gamma^5) \nu - g_\phi M_\phi |\chi|^2 \phi - \frac{1}{2} M_\phi^2 \phi^2.
    \label{t-channel UV}
\end{equation}
Integrating out the mediator $\phi$ for a low-energy DM scattering yields quadratic couplings of $G_s \simeq y_s g_\phi/ M_\phi$ and $G_p \simeq y_p g_\phi / M_\phi$. 

For completeness, we briefly comment on the UV model for the Yukawa interaction in  Eq.~\eqref{s-channel UV} and Eq.~\eqref{t-channel UV} to underline a possible probe in the SM channel. For $s$-channel mediator, DM can be introduced as an inert scalar singlet $\chi$, in which the Yukawa interaction is generated from an effective operator that couples scalar DM $\chi$ with Higgs $H$ and a lepton doublet $L$; \textit{i.e.}, $\frac{\Bar{L} H \chi N_R}{\Lambda}$. Alternatively, if DM is realized as a neutral component of an inert scalar doublet $\Phi_\chi$ under $SU(2)_L$, an interaction term $y \Bar{L} \Phi_\chi N_R$ suffices. We note that the singlet requires a Higgs to generate the Yukawa structure, whereas the doublet does not. For $t-$channel mediator, the interaction $\phi \bar{\nu}\nu$ can be generated with $\bar{L^c} i \tau_2 \Delta L$, where $\Delta$ is the Higgs triplet with its neutral component $\phi$.

\begin{figure}[htbp]
\centering
\includegraphics[width=0.8\linewidth]{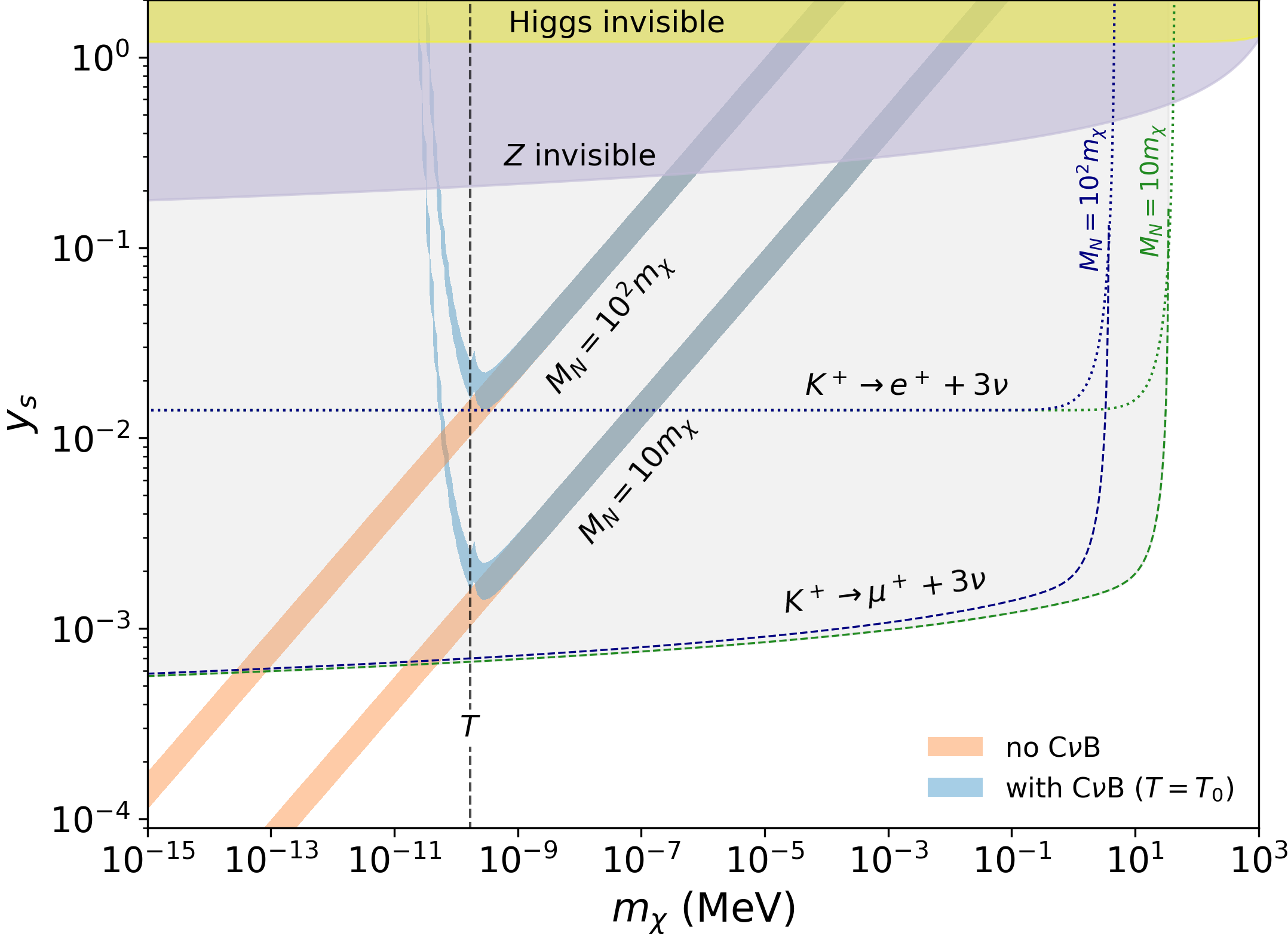}
\caption[Short-form caption]{The preferred region of parameter space for addressing the core-cusp problem in the full theory with an s-channel mediator for $M_N = 10m_\chi$ and $M_N = 10^2 m_\chi$ with $T = T_0$. Larger $M_N/m_\chi$ suppresses the interaction, shifting the preferred region up toward larger $y_\text{s}$. Existing upper bounds on the Yukawa coupling $y_s$ are set from the invisible decays of the Higgs boson (yellow band) and the Z boson (purple band). Additionally, bounds from the leptonic decay of the Kaon through $K^+ \rightarrow l^+ \chi \bar{N}$ where $l^+ = e^+,\mu^+$ are shown for $M_N = 10m_\chi$ (green curve) and $M_N = 10^2m_\chi$ (blue curve). All bounds are at 95\% confidence level. The flavor constraints rules out the entire preferred region of parameter space when the C$\nu$B is included. This is due to the screening effect that diminishes DM scattering strength.
}
\label{fig: full theory s-channel parameter space}
\end{figure}

\begin{figure}[htbp]
\centering
\includegraphics[width=0.8\linewidth]{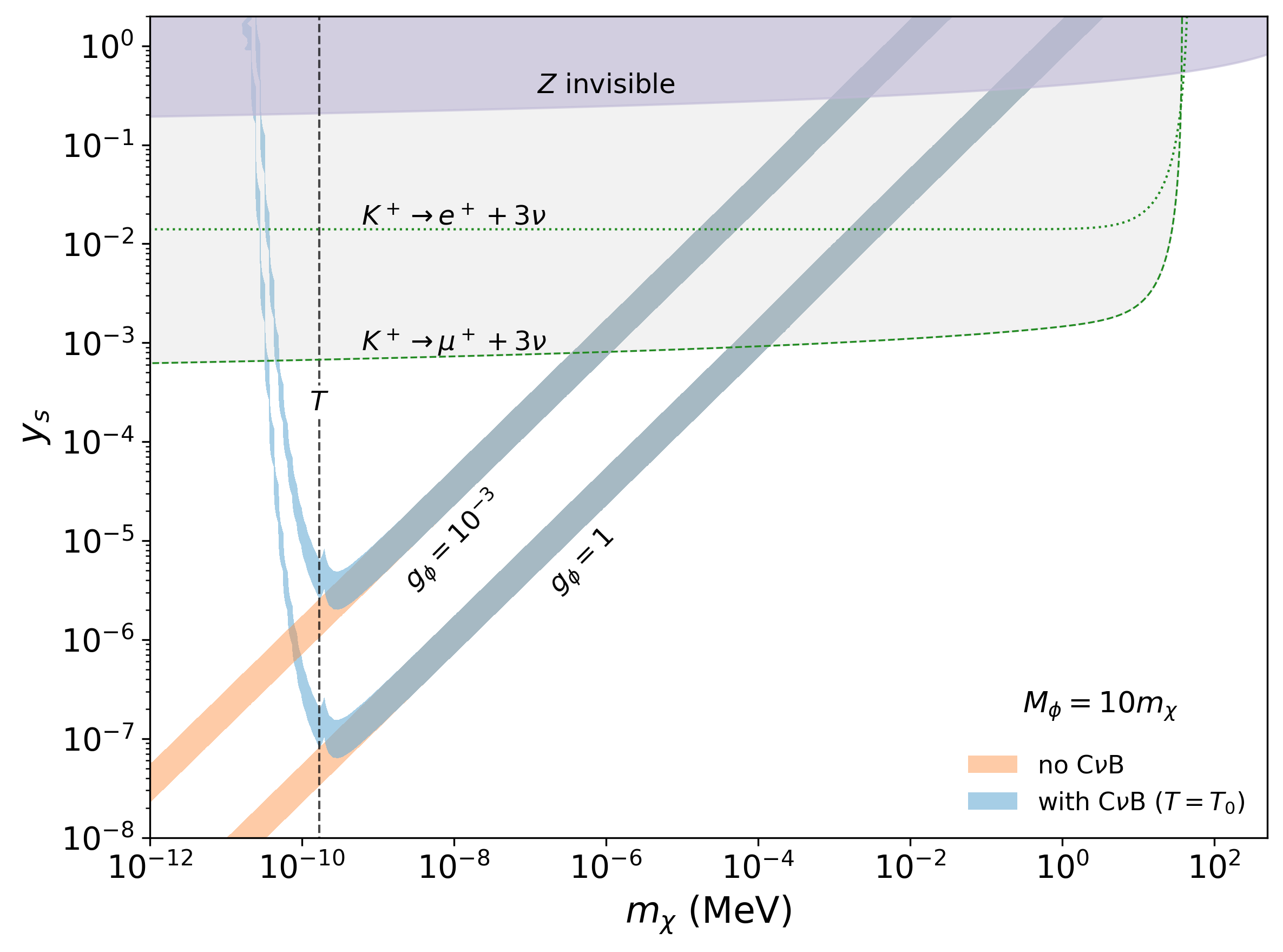}

\caption[Short-form caption]{The preferred region of parameter space for addressing the core-cusp problem in the full theory with a t-channel mediator for $g_\phi = 1$ and $g_\phi = 10^{-3}$ with $M_\phi = 10m_\chi$ and $T = T_0$. Larger $M_\phi/m_\chi$ and smaller coupling $g_\phi$ suppress the interaction, shifting the preferred region up toward larger $y_\text{s}$. The $y_s$ upper bound is set by the Z boson invisible decay (purple band). The leptonic decay of the Kaon bounds through $K^+ \rightarrow l^+ \phi \bar{\nu}$ where $l^+ = e^+,\mu^+$ are shown (green curve). The C$\nu$B effect shifts the allowed parameter space towards larger couplings in the region $m_{\chi} \lesssim T$, where it is excluded by the flavor constraint.
}
\label{fig: full theory t-channel parameter space}
\end{figure}

To chart the allowed parameter space in terms of coupling and DM mass, for simplicity, we assume a CP-conserving theory such that the pseudoscalar coupling $y_p$ vanishes. This allows us to focus on the coupling $y_s$, which precision measurements of SM particles can probe. These laboratory constraints have been studied in the context of fermionic DM (See ref.~\cite{Orlofsky_2021,Zhang_2024}). Following these works, we apply them here to our scalar DM-neutrino model. Moreover, we assume that we only account for the potential generated by Eq.~\eqref{Effective DM Interaction}. Other short-range terms, such as the quadratic interaction between the scalar (i.e. $\abs{\chi}^2 \abs{H}^2$), are ignored. Thus, the allowed regions should be interpreted as upper bounds. Furthermore, for the $t$-channel mediator, we consider the case in which $M_\phi \gg m_\chi$ such that $\phi$ is not the relevant degree of freedom.  figure~\ref{fig: full theory s-channel parameter space} and~\ref{fig: full theory t-channel parameter space} show the resulting parameter space for the $s$-channel and the $t$-channel, respectively.

The dominant constraints on $y_s$ arise from SM invisible decays. The interaction contributes to the invisible $Z$ decay via $Z \rightarrow \bar{\nu} \chi N$ and $Z \rightarrow \bar{\nu} \nu \phi$ for the $s$-channel and $t$-channel, respectively. The upper bound is set by the uncertainty in addition to its invisible decay width $\Gamma(Z \rightarrow \text{invisible}) \simeq 506\pm2$ MeV~\cite{ATLAS:2023ynf}, which is shown by the purple shaded area in figure~\ref{fig: full theory s-channel parameter space} and~\ref{fig: full theory t-channel parameter space}. If the coupling arises from a higher-dimensional operator such as $\Bar{L} H \chi N_R/\Lambda$, it also leads to the Higgs' invisible decay via $h \rightarrow \Bar{\nu} \chi N$. The upper bound is obtained by demanding that $\text{BR}(h\to\text{invisible}) < 24\% $~\cite{ATLAS:2019cid,CMS:2022qva} and is illustrated as a yellow shaded area in figure~\ref{fig: full theory s-channel parameter space}. However, this is already excluded by the $Z$ bound.

The SM invisible decay constraints discussed above are independent of the neutrino flavors. On the other hand, if DM $\chi$ couples to the neutrino flavor $\nu_\mu$ or $\nu_e$, there are additional flavor-dependent constraints from the kaon leptonic decay channel via $K^+ \rightarrow l^+ \chi \Bar{N}$ and $K^+ \rightarrow l^+ \phi \nu$ for the $s$-channel and $t$-channel, respectively, where $l^+ = e^+, \mu^+$. The corresponding decay rate is given by
    \begin{align}
    \Gamma_{K^+ \rightarrow l^+ \Phi \mathcal{N}} 
    &= \frac{y_s^2 G_F^2 m_K^3 F_K^2}{256 \pi^3} 
       \int_{(1+b)^2 z}^{(1 - \sqrt{y})^2} dx \,
       \Bigl[\,x + y - (x-y)^{2}\Bigr]\;
       \Bigl[(b^{2}-1)z + x\Bigr] \notag \\
    &\hspace{-1.5em} \times
    \frac{ 
    \sqrt{\,x^{2} + (1+b^{4})\,z^{2} - 2x(1+b^{2})z - 2b^{2}z^{2}\,}\;
    \sqrt{\,x^{2} - 2x(y+1) + (y-1)^{2}\,}
    }{x^{3}} \,, 
    \end{align}
where $\Phi \in \{\chi,\phi\}$ and $\mathcal {N} \in \{\bar{N},\nu\}$. The parameters are given as followed: $y = m_l^2 / m_K^2$, $z = m_\Phi^2/m^2_K$, $b = m_\mathcal{N}/m_\Phi$, and $F_K$ is the Kaon decay constant. Recasting the bounds on \(\text{Br}(K^+ \rightarrow e^+ + 3\nu) < 6 \times 10^{-5}\)~\cite{HEINTZE1979365} and \(\text{Br}(K^+ \rightarrow \mu^+ + 3\nu) < 1.0 \times 10^{-6}\)~\cite{Cortina_Gil_2021} yields the upper limit curves for $M_N = 10m_\chi$ (green curve) and $M_N = 10^2  m_\chi $ (blue curve) in figure~\ref{fig: full theory s-channel parameter space} for the $s$-channel mediator and the upperbound (green curve) in figure~\ref{fig: full theory t-channel parameter space} for the t-channel mediator. 

\newpage 

\section{Conclusion}
\label{sec: conclusion}

In this work, we investigated how corrections from the cosmic neutrino background modify the phenomenology of self-interacting scalar dark matter mediated by neutrino forces.  
We demonstrated that the impact depends on the hierarchy of mass scales $(m_\nu, m_\chi, T )$. In particular, the regimes $m_\nu \lesssim m_\chi \lesssim T$ are significantly affected since the repulsive background potential $V_{\text{bkg}}$ completely screens out the attractive vacuum potential $V_{\text{vac}}$.

We began by proposing an effective DM-neutrino interaction, as described in Eq.~\eqref{Effective DM Interaction}, for both scalar and pseudoscalar interactions. Following the formalism in ref.~\cite{Ghosh_2023}, we derived both the vacuum and background potentials for these interactions. A key feature is that the vacuum potential $V_{\rm vac}$ is attractive, whereas the background potential $V_{\rm bkg}$ is repulsive at the range $T^{-1} \lesssim m_\chi^{-1} \ll r$. Thus, in the regime $m_\chi \lesssim T$, this leads to a completely screened potential in the relativistic limit ($m_\nu \ll T$), which renders DM scattering negligible.

We then analyzed how the C$\nu$B modifies DM-neutrino phenomenology. We first reviewed the computation to capture non-perturbative effects from multiple neutrino exchange for the self-scattering cross section and the Sommerfeld Enhancement factor by solving the Schrödinger equation. The resulting self-scattering cross-section follows the Born approximation with $\sigma \propto G^4$. Due to the singular nature of the neutrino potential ($\sim 1/r^5$), we regularized it by introducing a short-distance cutoff $R = m_\chi^{-1}$, implemented as a flat potential for $r \le R$. We identified the viable region of parameter space for the small-scale structure problem and investigated those affected by the C$\nu$B correction. In particular, the C$\nu$B background screening significantly affects the allowed region with DM masses $m_\chi \lesssim T$, shifting the allowed region toward larger couplings.

Conversely, the C$\nu$B impact is negligible for higher DM masses. The effect of relevant parameters is discussed as follows. For the scalar and pseudoscalar interactions, there is no distinction in DM phenomenology. The massive neutrino mass affects the scattering only in the region with $m_\chi \sim m_\nu$ due to the exponentially suppressed vacuum potential. Moreover, at higher temperatures, the cosmic neutrino background can significantly affect DM with masses up to $m_\chi \lesssim T$ by completely screening the potential. 

In parallel, we investigated the effect of C$\nu$B on the Sommerfeld enhancement factor S in DM annihilation in the non-relativistic regime ($G_s \le m_\chi^{-1}$). While the enhancement generally increases with a stronger coupling, the screening effect from the induced background potential eliminates the enhancement. Thus, C$\nu$B renders the Sommerfeld enhancement negligible for DM annihilation in the coupling ranges considered.

Finally, we discussed UV completions for the scalar DM scenario in which the effective operator in Eq.~\eqref{Effective DM Interaction} arises from integrating out either a heavy fermion ($s$-channel mediator) or a scalar ($t$-channel mediator) introduced in Eqs.~\eqref{s-channel UV} and \eqref{t-channel UV}, respectively. We derived corresponding effective couplings $G_s$ and $G_p$, along with discussing the nature of the scalar DM candidate. For the $s$-channel mediator, the allowed parameter space is ruled out entirely. This conclusion is based on the scans over the Yukawa coupling $y_s$ and DM mass $m_\chi$, combined with bounds from invisible $Z$ and Higgs boson decays and flavor-dependent leptonic kaon decays. In contrast, the parameter space in $t$-channel scenario with similar bounds remains viable at $m_{\chi} \gtrsim T$ due to the lack of screening effect in this region.

In summary, C$\nu$B corrections can substantially reshape the phenomenology of neutrino-mediated, self-interacting scalar DM. The effect is significant when $m_\chi \lesssim T$, where the background potential completely screens out the vacuum potentials. The scalar DM–neutrino framework, with C$\nu$B effects included, remains capable of addressing small-scale structure tensions such as the core–cusp problem while staying consistent with current experimental constraints over a broad range of DM masses.

\appendix

\section{Calculation of the massive neutrino vacuum potential}
\label{appendix: Calculation of the massive neutrino vacuum potential}

For the scalar interaction, the scattering amplitude of DM scattering via the neutrino force in the figure~\ref{fig: vac vs bkg diagram} is given as 
\begin{align}
    i\mathcal{M}^s_{\rm vac} &= - \left (-\frac
    {iG_s}{\sqrt{2}} \right)^2  \int \frac{d^4 k}{(2 \pi)^4} {\rm Tr}\left[ S^0_\nu(k)S^0_\nu(k+q)\right] \times 2 , \nonumber \\
    &= \frac{-iG_s^2}{8\pi^2} \Big[ (4m_\nu^2 - q^2) B_0 (q^2, m_\nu^2, m_\nu^2) + 2 A_0(m_\nu^2) \Big],
\end{align}

where $A_0$ and $B_0$ are the generic Passarino-Veltman integrals. The factor of $2$ is due to the exchange of two Majorana neutrino propagators in the loop. These expressions are given as
\begin{subequations}
\begin{align}
     & A_0 (m_\nu^2) = m_\nu^2 \left( \Delta_\epsilon + 1 - \ln \frac{m_\nu^2}{\mu^2}\right), \tag{\theequation a} \label{PV Integral A0} \\
    & B_0 (q^2, m_\nu^2, m_\nu^2) = \Delta_\epsilon - \int^1_0 dx \ln\left[ \frac{-x(1-x)q^2 + m_\nu^2}{\mu^2} \right], \tag{\theequation b} \label{PV Integral B0} 
\end{align}
\end{subequations}
where $\Delta_\epsilon \equiv 2/\epsilon - \gamma + \ln 4 \pi$ is the divergence term in which $\gamma \approx 0.577$. The variables $d = 4 - \epsilon$ from the dimensional regularization scheme. The $\mu$ is the arbitrary renormalization scale. The integral is evaluated using the Fourier Transform \eqref{fourier transform}. This can be shown as 
\begin{align*}
     V_{\rm vac}^s(\bold{r}) &= +\frac{iG_s^2}{144 \pi^4 m_\chi^2 r} \int^{\infty}_{-\infty} d \rho \rho \left[(4m_\nu^2+\rho^2) B_0 (q^2, m_\nu^2, m_\nu^2) + 2 A_0(m_\nu^2)\right] e^{i \rho r}, 
     \\&= -\frac{iG_s^2}{144 \pi^4 m_\chi^2 r}  \int^{\infty}_{-\infty} d \rho_i \; \rho_i \left\{(4m_\nu^2 - \rho_i^2) \int^1_0 dx \ln\left[ \frac{-x(1-x)\rho_i^2 + m_\nu^2}{\mu^2} \right] \right\} e^{-\rho_i r}. 
\end{align*}
where we have used the change of variable $\rho = \abs{\bold{q}}$; $q^2 \approx -\rho^2 < 0$ in the NR-limit. Additionally, we deform the contour by changing the variable $\rho = i \rho_i$ where $\rho_i$ is now the imaginary part of $\rho$. This adjusts to the second equality where we end up with the discontinuity of the logarithm term along the imaginary axis from $\rho_i = m_\nu/\sqrt{x(1-x)}$ to $\rho_i = +\infty$. Note that the $A_0$ does not have singularities or branch cuts along the imaginary line. Therefore, the divergence term does not contribute to the integral.

Evaluating the integral analytically, we get the neutrino force potential as a Bessel Function of the Second Kind $K_2$ given as
\begin{equation}
    V_{\rm vac}^s(\bold{r}) = -\frac{3 G_s^2 m_\nu^2}{16\pi^3 m_\chi^2 r^3} K_2(2 m_\nu r).
\end{equation}
The asymptotic solutions of the scalar interaction at both short range ($r \ll m_\nu^{-1}$) and long range ($r \gg m_\nu^{-1}$) are given by
\begin{subequations}
    \begin{align}
    V_{\rm vac}^s(r \ll m_\nu^{-1})&\simeq -\frac{3G_s^2}{32\pi^3 m_\chi^2 r^5},
    \tag{\theequation a} \label{short range of vacuum potential of scalar}\\
    V_{\rm vac}^s(r \gg m_\nu^{-1}) &\simeq  -\frac{3 G_s^2}{64 m_\chi^2 } \left({\frac{m_\nu^3}{\pi^5 r^7}}\right)^{1/2} e^{-2m_\nu r}.
    \tag{\theequation b} \label{long range of vacuum potential of scalar}
    \end{align}
\end{subequations}
Following the same derivation, the neutrino potential from the pseudoscalar interaction is given as
\begin{equation}
    V^p_{\rm vac}(\bold{r}) = -\frac{G_p^2 m_\nu^2}{64\pi^3 m_\chi^2 r} \int^{\infty}_{2 m_\nu} d \rho_i \; \rho_i^3 e^{-\rho_i r} \left( 1 - \frac{4m_\nu^2}{\rho_i^2} \right)^{1/2}. 
\end{equation}
This neutrino force potential must be evaluated numerically. However, the asymptotic solutions of the pseudoscalar interaction at both short range ($r \ll m_\nu^{-1}$) and long range ($r \gg m_\nu^{-1}$) are given by
\begin{subequations}
    \begin{align}
    V_{\rm vac}^p(r \ll m_\nu^{-1})&\simeq -\frac{3G_p^2}{32\pi^3 m_\chi^2 r^5},\tag{\theequation a} \label{short range of vacuum potential of pseudoscalar}\\
    V_{\rm vac}^p(r \gg m_\nu^{-1}) &\simeq 
    - \frac{G_p^2}{8 m_\chi^2} \left({\frac{m_\nu}{\pi r}}\right)^{5/2} e^{-2m_\nu r}.
    \tag{\theequation b} \label{long range of vacuum potential of pseudoscalar}
    \end{align}
\end{subequations}

\section{Calculation of the massive neutrino background potential}
\label{appendix: Calculation of the massive neutrino background potential}
In this section, we follow the same derivation of the background potential as in ref. \cite{Ghosh_2023} for our proposed scalar and pseudoscalar interactions. For the scalar interaction, the background scattering amplitude is given as
\begin{equation*}
    i\mathcal{M}_{\rm bkg}^s = - \left(\frac{iG_s}{\sqrt{2}} \right)^2 \int \frac{d^4 k}{(2 \pi)^4} {\rm Tr}\left[ S^0_\nu(k)S^{\rm bkg}_\nu(k+q) + S^{\rm bkg}_\nu(k)S^0_\nu(k+q) \right] \times 2,
\end{equation*}
where the factor of $2$ is due to the exchange of two Majorana neutrino propagators in the loop. We adjust the form further by shifting the momentum from $k \rightarrow k - q$ in the second term and taking the delta function property $\delta(k^2 - m_\nu^2) = \frac{1}{2E_k}\left[ \delta(k^0 - E_k) + \delta(k^0 + E_k) \right]$ where $E_k = \sqrt{\bold{k}^2 + m_\nu^2}$. Applying the NR approximation $q \simeq (0, \textbf{q})$, we get
\begin{equation}
     \mathcal{M}_{\rm bkg}^s = -4 G_s^2 \int \frac{d^3k}{(2\pi)^3} \frac{n_+(\textbf{k}) + n_-(\textbf{k})}{2E_k} \left[ \frac{\textbf{k} \cdot \textbf{q} + 2m_\nu^2}{2\textbf{k} \cdot \textbf{q} - \textbf{q}^2} + \frac{\textbf{k} \cdot \textbf{q} - 2m_\nu^2}{2\textbf{k} \cdot \textbf{q} + \textbf{q}^2}  \right],
     \label{General Background Amplitude for scalar}
\end{equation}

Given the isotropic distribution, we have $n_{\pm}(\textbf{k}) = n_{\pm}(\kappa)$ where $\kappa \equiv \abs{k}$. We transform the integration into a spherical coordinate in which $\rho \equiv \abs{\textbf{q}}, \varepsilon \equiv \cos \theta; \textbf{k} \cdot \textbf{q} = \abs{k} \abs{q} \cos \theta$. The Amplitude \eqref{General Background Amplitude for scalar} can be written as
\begin{equation}
    \mathcal{M}_{\rm bkg}^s = +\frac{G_s^2}{2 \pi^2} \int^{+\infty}_0 d\kappa \frac{\kappa^2}{\sqrt{\kappa^2 + m_\nu^2}} [n_+(\kappa) + n_-(\kappa)] \int^1_{-1} d \varepsilon \frac{4m_\nu^2 + 4 \kappa^2 \varepsilon^2}{\rho^2 - 4 \kappa^2 \varepsilon^2}.
    \label{Isotropic Background Amplitude for scalar}
\end{equation}
The background potential can be determined by performing the Fourier transform of the non-relativistic amplitude. We get
\begin{equation}
    V_{\rm bkg}^s(r) = +\frac{G_s^2}{16\pi^3 m^2_\chi r^4} \int_0^{\infty} d\kappa \; \kappa \frac{n_+(\kappa) + n_-(\kappa)}{\sqrt{\kappa^2 + m_\nu^2}}  \left[1- 2\left(\kappa^2+m_\nu^2\right)r^2]  \sin (2 \kappa r)-2 \kappa r \cos (2 \kappa r) \right].
    \label{Isotropic Background Potential for scalar appendix}
\end{equation}

\noindent Following the same procedure, the isotropic background potential for the pseudoscalar interaction is given as
\begin{equation}
    V_{\rm bkg}^p(r) = +\frac{G_p^2}{16\pi^3 m^2_\chi r^4} \int_0^{\infty} d\kappa \; \kappa \frac{n_+(\kappa) + n_-(\kappa)}{\sqrt{\kappa^2 + m_\nu^2}}  \left [ (1- 2 \kappa^2 r^2)  \sin (2 \kappa r)-2 \kappa r \cos (2 \kappa r) \right ]. 
    \label{Isotropic Background Potential for pseudoscalar appendix}
\end{equation}

\newpage

\acknowledgments
The authors would like to thank S. Fichet for insightful discussions and for directing us to relevant literature that strengthened this study. The work of PU is supported in part
by the National Science, Research and Innovation Fund (NSRF) via Srinakharinwirot University under grant number 055/2569. PI and PU also acknowledge the National Science and Technology Development Agency, National e-Science Infrastructure Consortium, Chulalongkorn University, and the Chulalongkorn Academic Advancement into Its 2nd Century Project (Thailand) for providing computing infrastructure that has contributed to the results reported within this paper. CP is supported by Fundamental Fund 2568 of Khon Kaen University and Research Grant for New Scholar, Office of the Permanent Secretary, Ministry
of Higher Education, Science, Research and Innovation under contract no. RGNS 64-043. MS is supported by the Honours Program, Faculty of Science, Chulalongkorn University.
NT is supported by the NSRF via the Program Management Unit for Human Resources \& Institutional Development, Research and Innovation [grant number B13F670063]. \\




\bibliography{biblio}
\bibliographystyle{JHEP}

\end{document}